\documentclass[aps,pra,reprint,superscriptaddress]{revtex4-1}
\usepackage{amsmath}
\usepackage{amssymb}
\usepackage{mathrsfs}
\usepackage{graphicx}
\usepackage[caption=false]{subfig}
\usepackage{enumitem}
\usepackage{color}
\usepackage[percent]{overpic}
\usepackage{bm}
\usepackage{rotating}
\usepackage{hyperref}
\usepackage{cancel}
\usepackage{verbatim}
\usepackage{mathtools}
\usepackage{graphicx}
\usepackage{tensor}
\usepackage{verbatim}
\usepackage{MnSymbol}
\usepackage{booktabs}
\graphicspath{ {images/} }

\usepackage{braket}

\newcommand{\ii}{\mathrm{i}}

\renewcommand{\d}{\mathrm{d}}
\renewcommand{\vec}{\text{vec}}

\newcommand{\be}{\begin{equation}}
\newcommand{\bel}[1]{\begin{equation}\label{#1}}
\newcommand{\ee}{\end{equation}}

\newcommand{\tcr}{\textcolor{red}}

\begin{document}
\title{A classification of open Gaussian dynamics}

\author{Daniel Grimmer}
\email{dgrimmer@uwaterloo.ca}
\affiliation{Institute for Quantum Computing, University of Waterloo, Waterloo, ON, N2L 3G1, Canada}
\affiliation{Dept. of Physics and Astronomy, University of Waterloo, Waterloo, ON, N2L 3G1, Canada}

\author{Eric Brown}
\email{ericgb86@gmail.com}
\affiliation{ICFO-Institut de Ciencies Fotoniques, The Barcelona Institute of Science and Technology, 08860 Castelldefels (Barcelona), Spain}

\author{Achim Kempf}
\affiliation{Dept. of Applied Mathematics, University of Waterloo, Waterloo, ON, N2L 3G1, Canada}
\affiliation{Dept. of Physics and Astronomy, University of Waterloo, Waterloo, ON, N2L 3G1, Canada}
\affiliation{Institute for Quantum Computing, University of Waterloo, Waterloo, ON, N2L 3G1, Canada}
\affiliation{Perimeter Institute for Theoretical Physics, Waterloo, ON, N2L 2Y5, Canada}

\author{Robert B. Mann}
\email{rbmann@uwaterloo.ca}
\affiliation{Dept. of Physics and Astronomy, University of Waterloo, Waterloo, ON, N2L 3G1, Canada}
\affiliation{Institute for Quantum Computing, University of Waterloo, Waterloo, ON, N2L 3G1, Canada}
\affiliation{Perimeter Institute for Theoretical Physics, Waterloo, ON, N2L 2Y5, Canada}

\author{Eduardo Mart\'{i}n-Mart\'{i}nez}
\email{emartinmartinez@uwaterloo.ca}
\affiliation{Institute for Quantum Computing, University of Waterloo, Waterloo, ON, N2L 3G1, Canada}
\affiliation{Dept. of Applied Mathematics, University of Waterloo, Waterloo, ON, N2L 3G1, Canada}
\affiliation{Perimeter Institute for Theoretical Physics, Waterloo, ON, N2L 2Y5, Canada}

\begin{abstract}
We introduce a classification scheme for the generators of bosonic open Gaussian dynamics, providing instructive diagrams description for each type of dynamics. Using this classification, we discuss the consequences of imposing complete positivity on Gaussian dynamics. In particular, we show that non-symplectic operations must be active to allow for complete positivity. In addition, non-symplectic operations can, in fact, conserve the volume of phase space only if the restriction of complete positivity is lifted. We then discuss the implications for the relationship between information and energy flows in open quantum mechanics.
\end{abstract}

\maketitle
\section{Introduction}
The study of quantum systems that interact with an environment, i.e., the study of open quantum dynamics, is important for a wide variety of reasons. It is important, for example, for quantum thermodynamical studies into how quantum systems equilibrate with heat baths. The study of open quantum systems is also particularly important for experiments in which environment-induced decoherence and dephasing is to be either minimized or exploited \cite{Nielsen:2000}. 

In order to solve quantum dynamics, a particularly powerful tool is the formalism of Gaussian quantum mechanics (GQM), in that it allows for a large (or even infinite) decrease in the overhead for describing quantum states and their transformations. There are two conditions for the GQM formalism to be applicable, namely that the states are representable by Wigner functions that are Gaussian and that the transformations in question preserve this Gaussianity. Fortunately, these two conditions are met by many of the phenomena which are experimentally accessible and relevant,  and Gaussian approaches to quantum information and resource theory have been well established \cite{weedbrook, adesso1, lami}.

In the present paper, we will bring to bear the tools of Gaussian quantum dynamics to the study of the dynamics of open quantum systems that obey these Gaussianity conditions. In particular, beginning from the most general form of the master equation of such a system, we categorize the dynamic's generators according to several criteria, namely, whether the dynamics allows for the flow of energy or quantum information between the system and its environment, whether the effect of the dynamics is state-dependent and whether it mixes different modes. Following this categorization, we will analyze the complete positivity of each contribution to the dynamics as well as their effect on the volume of phase space.  This analysis adds an important theoretical tool to the growing literature regarding Gaussian open dynamics \cite{koga, nicacio, nicacio2}.

While, in the literature, many of these questions have been asked before, we here develop a unified analysis of open Gaussian evolution in the Heisenberg picture. This unified analysis then allows us to comprehensively categorize the various types of contributions to the master equation and their impact on the open dynamics.  

In Sec. \ref{ReviewGQM}, we first give a brief review of GQM, establishing our notation. In Sec. \ref{Characterizing}, we describe four possible types of behavior that open Gaussian dynamics may display and track their origins to terms in the master equation.  Then we explicitly partition the dynamics into various parts according to the phenomena they support. To aid in understanding each type of dynamics, in Sec. \ref{Naming}, we produce phase space plots for each type of dynamics given by the partition and analyze their effect on a generic Gaussian state.

\section{A brief review of Gaussian Quantum Mechanics}\label{ReviewGQM}

\subsection{Symplectic Formalism}
Let us consider a system of $N$ coupled bosonic modes (e.g., harmonic oscillators), with the $n^\text{th}$ mode fully characterized by its creation and annihilation operators, \mbox{$\hat{a}_n^\dagger$ and $\hat{a}_n$,} which obey the canonical Bosonic commutation relations,
\begin{align}\label{CanonicalCommsAA}
[\hat{a}_n,\hat{a}_m]
=[\hat{a}_n^\dagger,\hat{a}_m^\dagger]
=0
\quad\text{and}\quad
[\hat{a}_n,\hat{a}_m^\dagger]
=\delta_{nm} \, \hat{\openone}
\end{align}
where  $\delta_{nm}$ is the Kronecker delta and $\hat{\openone}$ is the identity operator on the system's Hilbert space.

One can equivalently characterize the system in terms of its quadrature operators, \mbox{$\hat{q}_n$ and $\hat{p}_n$,} which are constructed from the creation and annihilation operators as
\bel{qpDef}
\hat{q}_n=\frac{1}{\sqrt{2}}(\hat{a}_n^\dagger+\hat{a}_n)
\quad\text{and}\quad
\hat{p}_n=\frac{\ii}{\sqrt{2}}(\hat{a}_n^\dagger-\hat{a}_n).
\ee
From \eqref{CanonicalCommsAA}, these quadrature operators obey their own canonical commutation relations, 
\begin{align}\label{CanonicalCommsQP}
[\hat{q}_n,\hat{q}_m]
=[\hat{p}_n,\hat{p}_m]
=0
\quad\text{and}\quad
[\hat{q}_n,\hat{p}_m]
=\ii \, \delta_{nm} \, \hat{\openone}.
\end{align}
It is convenient to collect these $2N$ observables in the following operator-valued phase space vector,
\bel{XhatDef}
\hat{\bm{X}}
\coloneqq
(\hat{q}_1,\hat{p}_1,\hat{q}_2,\hat{p}_2,\dots,\hat{q}_N,\hat{p}_N)^\intercal.
\ee
Throughout this paper the index $n$ will run from $1$ to $N$, labeling the bosonic modes. The index $m$ will run from 0 to $\infty$. All other Latin indices ($i,\, j,\, k$, etc.) will run from $1$ to $2N$ labeling the components of a vector in phase space. For such phase space vectors, we will use subscripts for the components of column vectors and superscripts for the components of row vectors. Additionally we adopt Einstein's summation notation for repeated indices. 

Note that every pair of these observables, say $\hat{X}_j$ and $\hat{X}^k$, commute to a (potentially zero) multiple of the identity operator on the system's Hilbert space. Thus their commutation relations are completely captured by the phase space matrix, $\Omega$, defined by
\bel{OmegaDef}
[\hat{X}_j,\hat{X}^k]
=\ii \ \Omega_j{}^k \, \hat{\openone}.
\ee
Here, $\Omega$ is a \mbox{$2N$ by $2N$} real-valued, antisymmetric, invertible matrix with  \mbox{$\Omega^{-1}=\Omega^T=-\Omega$}, which means that it represents a symplectic form. This symplectic form is the key to allowing one to translate the description of the quantum dynamics of the system from Hilbert space to a phase space formalism. 

Concretely, for the operator ordering given by \eqref{XhatDef} the symplectic form is given by 
\bel{OmegaExplicit}
\Omega
=\bigoplus_{n=1}^N \omega
=\openone_N\otimes\omega; \ \ \ \ \omega
=\begin{pmatrix}
0 & 1\\
-1 & 0
\end{pmatrix}
\ee
where $\openone_N$ is $N$-dimensional identity operator.

We note that an alternate operator ordering,
\bel{AltXhatDef}
\hat{\bm{X}}_\text{alt}=(\hat{q}_1,\dots\hat{q}_N,\hat{p}_1,\dots,\hat{p}_N)^\intercal
\ee
is also common in the literature and would yield an alternate expression for the symplectic form,
\bel{AltOmegaExplicit}
\Omega_\text{alt}
=\omega\otimes\openone_N
=\begin{pmatrix}
0 & \openone_N\\
-\openone_N & 0
\end{pmatrix}.
\ee
We prefer the ordering given by \eqref{XhatDef} as it has the conjugate pairs of observables adjacent. This ordering is helpful in addressing individual modes and in characterizing dynamics as either single-mode or multi-mode (as we will see in Sec. \ref{Characterizing}).

\subsection{Gaussian States}
Having captured the algebraic structure of our system's Hilbert space in terms of the phase space matrix, $\Omega$, we now discuss a class of quantum states which can be described in terms of simple phase space objects, namely Gaussian states.

In order to do this we note that Quantum Mechanics can be fully formulated in terms of pseudo-probability distributions on phase space \cite{Groenewold,Moyal}. In particular, a state represented in Hilbert space by a density matrix $\rho$ is equivalently represented by the following Wigner pseudo-probability distribution \cite{Wigner}:
\be
W(\bm{q},\bm{p})=\frac{1}{\pi^N}\int_{-\infty}^\infty \d^N \bm{s}
\bra{\bm{q}+\bm{s}}\rho\ket{\bm{q}-\bm{s}}\exp(-2\ii \, \bm{p}\cdot\bm{s})
\ee
where $\bm{p}\cdot\bm{s}$ is the standard inner product in $\mathbb{R}^N$.

Gaussian states are defined as quantum states that have Gaussian Wigner functions, that is Wigner functions of the form,
\be
W(\bm{Y})=\frac{1}{\pi^N\sqrt{\text{det}(\sigma)}}
\exp\big(-
(\bm{Y}-\bm{X})^\intercal
\sigma^{-1}
(\bm{Y}-\bm{X})
\big)
\ee
where $\bm{X}$ is a real-valued vector of length $2N$ and that $\sigma$ is a real-valued $2N$ by $2N$ symmetric matrix. The vector $\bm{X}$ captures the system's first statistical moments, i.e., mean of each observable,
\be
\bm{X}\coloneqq\langle\hat{\bm{X}}\rangle.
\ee
The matrix $\sigma$ captures the system's second statistical moments, i.e., the covariance between each pair of observables,
\bel{Vdef}
\sigma_j{}^k
\coloneqq
\big\langle
\hat{X}_j \, \hat{X}^k
+ \hat{X}^k \, \hat{X}_j
\big\rangle
-2\big\langle\hat{X}_j\big\rangle
\big\langle\hat{X}^k\big\rangle.
\ee

We note that an alternate definition for the covariance matrix, \mbox{$\sigma_\text{alt}=\sigma/2$} is common. We prefer the notation defined by \eqref{Vdef} as it removes many factors of two from the formalism (knowing this may be helpful when comparing our results with other literature). 

Just as a valid density matrix is a normalized positive semi-definite self-adjoint operator, the corresponding restriction on the Gaussian states is  \cite{Simon1994}
\bel{SigmaPosCond}
\sigma\geq\ii \, \Omega,
\ee
with pure states saturating the inequality. Throughout this text, the notation \mbox{$P\geq Q$} is used to mean that \mbox{$P-Q$} is positive semi-definite (i.e., in our notation \mbox{$P-Q\geq0$}). A matrix $P$ is said to saturate the inequality \mbox{$P\geq Q$} if \mbox{$P-Q$} has at least one vanishing eigenvalue, putting \mbox{$P-Q$} on the boundary of the positive semi-definite cone.

The condition \eqref{SigmaPosCond} implies that the covariance matrix is positive semi-definite by the following argument. Since conjugating the entries of a positive semi-definite matrix keeps it positive semi-definite, we have
\be
\sigma\geq\ii \, \Omega \Rightarrow 
\sigma=\sigma^*\geq(\ii \, \Omega)^*
=-\ii \, \Omega
\ee
where we have used the fact that $\sigma$ and $\Omega$ are both real-valued. Since $\sigma$ is greater than both $\ii \, \Omega$ and $-\ii \, \Omega$, it is greater than their average as well, thus $\sigma\geq0$. This implication cannot be reversed, in addition to keeping the covariance matrix positive semi-definite Eq. \eqref{SigmaPosCond} also enforces the the uncertainty principle by preventing the covariances from being arbitrarily small.

It is useful to visualize Gaussian states as hyperellipsoids in phases space corresponding to the all the points within one deviation of $\bm{X}$ with respects to $\sigma$. Specifically, in the case of a single mode ($N=1$), one can think of a Gaussian state as an ellipse in a 2-dimensional phase space centered at $\bm{X}=(\langle\hat{q}\rangle,\langle\hat{p}\rangle)$ and with major and minor axes orientation depending on the eigensystem of $\sigma$. In Sec. \ref{Naming} we present several figures using this visualization technique. 

Many relevant states for both theory and experiment are Gaussian states. For instance, taking a single mode ($N=1$) to be a harmonic oscillator, its thermal states (with respects to its free Hamiltonian) are described by
\bel{ThermalStateDef}
\bm{X}=0
\quad \text{and} \quad
\sigma=\begin{pmatrix}
\nu & 0 \\
0 & \nu \\
\end{pmatrix}
\ee
where $\nu\geq1$ is a monotone function of the temperature. In the ellipsoid picture, described above, this state corresponds to a circle of radius $\sqrt{\nu}$ centered at $\bm{X}=0$.

Pure coherent states are described by
\bel{CoherentStateDef}
\bm{X}=\begin{pmatrix}
\langle \hat{q} \rangle \\
\langle \hat{p} \rangle \\
\end{pmatrix}
\quad \text{and} \quad
\sigma
=\begin{pmatrix}
1 & 0 \\
0 & 1 \\
\end{pmatrix}.
\ee
Visualized as an ellipsoid, this state corresponds to  a circle of unit radius centered at $\bm{X}$.

A family of single-mode squeezed states are described by
\bel{N1SqueezedStateDef}
\bm{X}=0
\quad \text{and} \quad
\sigma=\begin{pmatrix}
\sigma_{qq} & 0 \\
0 & \sigma_{pp} \\
\end{pmatrix}
\ee
obeying the uncertainty principle $\sigma_{qq} \, \sigma_{pp}\geq1$. This state corresponds to an ellipsoid centered at $\bm{X}=0$ with its major and minor axes in the $q$ and $p$ directions with lengths $\sqrt{\sigma_{qq}}$ and $\sqrt{\sigma_{pp}}$ respectively. More generally, any single-mode squeezed states is Gaussian.

\subsection{Unitary Gaussian Transformations}
With our algebraic structure and Gaussian states rewritten in terms of phase space objects, we now turn our attention toward the unitary transformations which preseve the Gaussianity of the states they act on. Such transformations are called Gaussian unitary transformations.

Gaussian unitaries are generated by Hamiltonians which are quadratic in the quadrature operators \cite{Schumaker1986}. Any such Hamiltonian can be converted into the standard form,
\bel{QuadHamForm}
\hat{H}=\frac{1}{2}\hat{\bm{X}}^\intercal  \, F \, \hat{\bm{X}}
+\bm{\alpha}^\intercal\hat{\bm{X}}
\ee
where $F$ is a real-valued $2N\times 2N$ symmetric matrix and $\bm{\alpha}$ is real-valued vector of length $2N$. In Appendix \ref{AppGQM} we demonstrate this fact with a general quadratic Hamiltonian.

In the Heisenberg picture, evolution under \eqref{QuadHamForm} yields,
\bel{HeisenbergEqsGaussian}
\frac{\d}{\d t}\bm{\hat{X}}
=\ii[\hat{H},\bm{\hat{X}}]
=\Omega (F \bm{\hat{X}}+\bm{\alpha} \, \hat{\openone}).
\ee
Note that $\hat{H}$ is a linear map on the system's Hilbert space and acts on $\bm{\hat{X}}$ componentwise. On the other hand, $F$ is a linear map on the system's phase space and acts on $\bm{\hat{X}}$ as a phase space vector, yielding linear combinations of its (operator-valued) components. We demonstrate this computation in Appendix \ref{AppGQM}. 

From \eqref{HeisenbergEqsGaussian}, the evolution of $\bm{X}$ and $\sigma$ under \eqref{QuadHamForm} can be computed as,
\begin{align}
\label{SymplecticDiffXUpHam}
\frac{\d}{\d t}\bm{X}
&=\Omega (F \bm{X}+\bm{\alpha})\\
\label{SymplecticDiffVUpHam}
\frac{\d}{\d t}\sigma
&=(\Omega F) \, \sigma
+\sigma \, (\Omega F)^\intercal.
\end{align}

For a time independent Hamiltonian, integrating \eqref{HeisenbergEqsGaussian} for a time interval $[0,t]$ yields, 
\bel{GaussianUnitaryUpdate}
\bm{\hat{X}}(t)
=\hat{U}_\text{G}(t)\hat{\bm{X}}(0)\hat{U}_\text{G}(t)^\dagger
=S(t)\hat{\bm{X}}(0)+\bm{d} \, \hat{\openone}
\ee
where $\hat{U}_\text{G}(t)=\exp(\ii \, \hat{H} \, t)$ and
\begin{align}
\label{SHamDef}
S(t)&=\text{exp}(\Omega F \, t)\\
\label{dHamDef}
\bm{d}(t)&=\frac{\text{exp}(\Omega F \, t)-\openone_{2N}}{\Omega F} \, \Omega\bm{\alpha}.
\end{align}
Again note that $\hat{U}_\text{G}(t)$ is a linear map on the system's Hilbert space and acts on $\bm{\hat{X}}$ componentwise. On the other hand, $S(t)$ is a linear map on the system's phase space and acts on $\bm{\hat{X}}$ as a phase space vector, yielding linear combinations of its (operator-valued) components. We demonstrate this computation in Appendix \ref{AppGQM}.

Finally note that $\Omega F$ does not need to be invertible to make sense of \eqref{dHamDef}, if one understands it in terms of the following definition:
\bel{(ExpX-1)byXDef}
\frac{\text{exp}(M \, t)-\openone}{M}
\coloneqq\sum_{m=0}^\infty \frac{t^{m+1}}{(m+1)!}M^m
\ee
for a general square matrix $M$.

More generally, any transformation of the form,
\bel{GeneralUnitaryUpdate}
\bm{\hat{X}}
\longrightarrow
S\hat{\bm{X}}+\bm{d} \, \hat{\openone}
\ee
with generic real-valued $S$ and $\bm{d}$ can be implemented by evolving under a (potentially time dependent) quadratic Hamiltonian as long as it preserves the commutation relation (i.e., the symplectic form). Computing the evolution of the system's commutation relations under \eqref{GeneralUnitaryUpdate} one finds the symplectic form is updated as
\be
\Omega
\longrightarrow
S\Omega S^\intercal.
\ee
A linear transformation that preserves the symplectic form, i.e., has 
\bel{SympTranDef}
S\Omega S^\intercal=\Omega,
\ee
is called a symplectic transformation. Thus Gaussian unitary transformations on Hilbert space implement symplectic-affine transformations on phase space.

The effect of such a symplectic-affine update on the mean vector and covariance matrix can be quickly computed to be
\begin{align}
\label{SymplecticXUp}
\bm{X}&\longrightarrow S \, \bm{X}+\bm{d}\\
\label{SymplecticVUp}
\sigma&\longrightarrow S \, \sigma \, S^\intercal.
\end{align}

\subsection{Gaussian Channels}\label{OpenGQM}
The unitary transformations described in the previous section are not the most general class of transformations that  preserve the Gaussian nature of the state.

Analogously to the Stinespring dilation theorem, any completely positive trace preserving transformation that preserves Gaussianity can be written as a symplectic-affine (Gaussian unitary) transformation in a larger phase space \cite{GaussianDilation}.

Given the form of the Gaussian unitary update, \eqref{SymplecticXUp} and \eqref{SymplecticVUp}, the most general form for an open Gaussian update of $\bm{X}$ and $\sigma$ is,
\begin{align}
\label{GeneralUpdateX}
\bm{X}&\to T\bm{X}+\bm{d}\\
\label{GeneralUpdateV}
\sigma&\to T \, \sigma \, T^\intercal+R
\end{align}
where $T$ and $R$ are real-valued $2N\times 2N$ matrices, $R$ is symmetric, and $\bm{d}$ is a real-valued vector of length $2N$. We demonstrate this in Appendix \ref{AppGQM}.

Generally, any transformation, $T$, $\bm{d}$, $R$, can be realized in this way as long as the complete positivity condition,
\bel{FiniteCPCond}
R+\ii \, \Omega-\ii \, T \, \Omega \,  T^\intercal\geq0
\ee
is obeyed \cite{GQMRev}. We sketch a proof of this fact in Appendix \ref{AppGQM}. Recall that the notation $P\geq 0$ here means that $P$ is positive semi-definite.

Let us consider a general differential update:
\begin{align}\label{GaussianDiffUpdate}
T&=\openone_{2N}+\d t \ \Omega  \, A\\
\bm{d}&=\d t \ \Omega \, \bm{b}\\
R&=\d t \ C
\end{align}
where $A$, and $C$ are real-valued $2N\times 2N$ matrices with $C$ symmetric, and $\bm{b}$ is a real-valued vector of length $2N$. Note that since $\Omega$ is invertible we are justified in our assumption that a factor of $\Omega$ can be pulled out of $A$ and $\bm{b}$. This will be helpful later in connecting these generators to the Hamiltonian which generates the unitary evolution in Hilbert space.

From \eqref{GeneralUpdateX} and \eqref{GeneralUpdateV} one finds the most general form for the Gaussian master equations,
\begin{align}
\label{GeneralDiffXUp}
\frac{\d}{\d t}\bm{X}(t)
&=\Omega(A\bm{X}(t)+\bm{b})\\
\label{GeneralDiffVUp}
\frac{\d}{\d t}\sigma(t)
&=(\Omega A) \, \sigma(t)
+\sigma(t) \, (\Omega A)^\intercal
+C.
\end{align}
The dynamical effect of the $A$ term is to implement rotations, squeezing, and amplifications in phase space, whereas the $\bm{b}$ and $C$ terms yield state-independent translation and noise respectively (as can be seen in Sec. \ref{Naming}).

From \eqref{FiniteCPCond} one finds that in order for this differential transformation to be completely positive one needs,
\bel{DiffCPCond}
C\geq\ii \Omega (A-A^\intercal)\Omega.
\ee
Recall that the notation $P\geq Q$ means that $P-Q$ is positive semi-definite.

For time independent generators, one can solve the Gaussian master equations analytically yielding $T$, $\bm{d}$ and $R$ in terms of $A$, $\bm{b}$, and $C$. Specifically one finds,
\begin{align}\label{TdRExplicit}
T(t)&=\exp(\Omega A  \, t)\\
\bm{d}(t)&=\frac{\exp(\Omega A \,  t)-\openone_{2N}}{\Omega A}\bm{b}\\
\label{RfromAbC}
R(t)&=\text{vec}^{-1}\Big(\frac{\exp((\Omega A\otimes\Omega A) \,  t)-\openone_{4N}}{\Omega A\otimes\Omega A} \ \text{vec}(C)\Big).
\end{align}
We demonstrate this calculation in Appendix \ref{AppGQM}. 

First note that in \eqref{RfromAbC} we have introduced the $\vec$ operation which turns a matrix into a vector. Specifically, it maps outer products to tensor products as
\bel{OuterToTensor}
\vec(\lambda \ \bm{u}\bm{v}^\intercal)
=\lambda \ \bm{u}\otimes\bm{v}
\ee
for some scalar $\lambda$ and vectors $\bm{u}$ and $\bm{v}$. Since any matrix can be expanded as a sum of outer products, one quickly finds that for any matrices $X$, $Y$ and $Z$
\bel{vecIdentity}
\vec(X \, Y \, Z^\intercal)=(X\otimes Z)\vec(Y).
\ee
This operation can be represented by the vector formed by taking the entries of a matrix in order as follows,
\be
\vec\begin{pmatrix}
a & b \\
c & d
\end{pmatrix}
=(a,b,c,d)^\intercal.
\ee
Note that $\text{vec}^{-1}$ is trivially defined by ``restacking'' the matrices entries.

Note that, as before, $\Omega A$ and $\Omega A\otimes\Omega A$ do not need to be invertible to make sense of these solutions, as they can be defined in terms of the series \eqref{(ExpX-1)byXDef}.

\section{Characterizing Gaussian Master Equations}\label{Characterizing}
In this section we partition the dynamics that the general Gaussian master equations, \eqref{GeneralDiffXUp} and \eqref{GeneralDiffVUp}, can produce. In particular we will classify Gaussian dynamics according to the following four dichotomies:
\begin{itemize}
    \item Symplectic vs. Unsymplectic
    \item Passive vs. Active
    \item Single-Mode vs. Multi-Mode
    \item State-Dependent vs. State-Independent.
\end{itemize}
In the next subsection we define each of these dichotomies and flesh out their physical relevance. Following this we explicitly partition Gaussian dynamics along these four dichotomies. Finally we analyze the partition looking at the complete positivity of each type of dynamics as well as its effect on the volume of phase space.

\subsection{Classification of Gaussian Evolution}\label{Criteria} 
\subsubsection{Symplectic vs. Unsymplectic}
As discussed above, Gaussian transformations of the form \eqref{SymplecticXUp} and \eqref{SymplecticVUp} which preserves the symplectic form (see Eq. \eqref{SympTranDef}) corresponds to a unitary transformation in Hilbert space. Such symplectic-affine transformations on phase space are here called \textit{symplectic}.

Examining the form of a general open Gaussian update, \eqref{GeneralUpdateX} and \eqref{GeneralUpdateV}, one sees that this channel is symplectic (i.e., unitary in Hibert space) if and only if $R$ vanishes and $T$ preserves the symplectic form as,
\bel{TSympCond}
R=0,
\quad\quad
T \, \Omega \,  T^\intercal=\Omega.
\ee
Any channel which does not meet these two conditions corresponds to a non-unitary transformation in Hilbert space. From now on we will call such transformations \textit{unsymplectic}.

Taking our transformation to be differential (as in Eq. \eqref{GaussianDiffUpdate}) the conditions for the dynamics to be symplectic, i.e., \eqref{TSympCond}, becomes
\bel{DiffSympCond}
C=0,
\quad\quad
A=A^\intercal.
\ee
Thus we can identify the presence of an antisymmetric part of $A$, 
\bel{AuDef}
A_\textsc{u}\coloneqq\frac{1}{2}\big(A-A^\intercal\big),
\ee
as well as any non-zero $C$ term as being responsible for any unsymplectic dynamics. We identify these as the unsymplectic parts of the dynamics (hence the subindex U in $A_\textsc{u}$ for unsymplectic). 

The other parts of the dynamics, the symmetric part of $A$,
\bel{AsDef}
A_\textsc{s}\coloneqq\frac{1}{2}\big(A+A^\intercal\big),
\ee
and any $\bm{b}$ term can thus be identified as the symplectic parts of the dynamics. Note that if \eqref{DiffSympCond} holds then the open dynamics given by \eqref{GeneralDiffXUp} and \eqref{GeneralDiffVUp} reduces to the Hamiltonian dynamics given by \eqref{SymplecticDiffXUpHam} and \eqref{SymplecticDiffVUpHam}. From this we can identify $F=A_\textsc{s}$ and $\bm{\alpha}=\bm{b}$. Thus the generators of the symplectic part of the dynamics are associated with the effective Hamiltonian
\begin{align}\label{HeffDef}
\hat{H}_\text{eff}
&\coloneqq\frac{1}{2}\hat{\bm{X}}^\intercal  \, A_\textsc{s} \, \hat{\bm{X}}
+\bm{b}^\intercal\hat{\bm{X}}\\
&\nonumber
=\frac{1}{4}\hat{\bm{X}}^\intercal  \, (A+A^\intercal) \, \hat{\bm{X}}
+\bm{b}^\intercal\hat{\bm{X}}.
\end{align}

\subsubsection{Passive vs. Active}
In addition to classifying whether the dynamics are symplectic or not, we can also characterize the dynamics by their effect on the average total excitation number,
\begin{align}
\langle \hat{n}\rangle
=\sum_{n=1}^N\langle q_n^2+p_n^2\rangle
-\frac{1}{2}.
\end{align}
Here dynamics are considered either active or passive depending on whether they change or maintain the average total excitation number respectively. In the case where all $N$ modes have the same fundamental energy scale $\nu$ this quantity is related to the system's average free energy as 
\be
\langle \hat{H}_0\rangle
=\nu\big(\langle\hat{n}\rangle+N/2\big).
\ee
The average total excitation number can be written in terms of $\sigma$ and $\bm{X}$ as,
\bel{H0VX}
\langle \hat{H}_0\rangle
=\text{Tr}(\sigma/2+\bm{X}\bm{X}^\intercal).
\ee
Note that the trace in this equation is over the system's phase space.

The rate of change of the expected free energy can be computed using \eqref{GeneralDiffXUp} and \eqref{GeneralDiffVUp} as
\begin{align}\label{Activity}
\frac{\d}{\d t}\langle \hat{H}_0\rangle
&=\text{Tr}\Big(\big(\Omega A+(\Omega A)^\intercal\big)\big(\sigma/2+\bm{X}\bm{X}^\intercal\big)\\
&+2(\Omega\bm{b})^\intercal\bm{X}
+\text{Tr}\big(C\big).
\end{align}
Therefore any change in the expected free energy must be attributed to at least one of the following conditions: either $\Omega A+(\Omega A)^\intercal\neq0$, or $\bm{b}\neq0$, or $\text{Tr}(C)\neq0$. We can thus identify the entirety of $\bm{b}$ as an active component of the dynamics. In the same fashion, the part of $A$ which is symmetric when multiplied by $\Omega$ on the left is also active. Explicitly this is given by
\begin{align}
A_\textsc{a}
&\coloneqq \frac{1}{2}(\Omega)^{-1}\big(\Omega A+(\Omega A)^\intercal\big)\nonumber \\
&=\frac{1}{2}\big(A+\Omega^{-1} A \Omega^\intercal\big)
\label{AaDef}
\end{align}
On the other hand the part of $A$ which is antisymmetric when multiplied by $\Omega$ on the left,
\begin{align}
A_\textsc{p}
&\coloneqq \frac{1}{2}(\Omega)^{-1}\big(\Omega A-(\Omega A)^\intercal\big)\nonumber \\
&=\frac{1}{2}\big(A-\Omega^{-1} A \Omega^\intercal \big)
\label{ApDef}
\end{align}
is passive.

The $C$ term will always be in total active: since $C$ is positive semi-definite, $C\neq0$, implies $\text{Tr}(C)>0$. However, different parts of $C$ can be considered either active or passive depending on their trace. At the moment (and we will revisit this later) there is not a natural way to decompose $C$ into a ``traceful'' and ``traceless'' part.

\subsubsection{State-Dependent vs State-Independent}
We can further classify dynamics as either being state-dependent or state-independent. Mathematically, we draw this distinction by considering which terms in equations \eqref{GeneralDiffXUp} and \eqref{GeneralDiffVUp} are coupled or decoupled from the instantaneous system state, $\bm{X}(t)$ and $\sigma(t)$. For instance in equation \eqref{GeneralDiffXUp} the $\Omega A \bm{X}(t)$ term depends on $\bm{X}(t)$ while the $\Omega\bm{b}$ term does not. Similarly in equation \eqref{GeneralDiffVUp} the $\Omega A \,  \sigma(t)$ and $\sigma(t) (\Omega A)^\intercal$ are state dependent whereas the $C$ term is not. In summary $C$ and $\bm{b}$ are state-independent terms, whereas the $A$ term is state-dependent.

\subsubsection{Single-Mode vs. Multi-Mode}
Finally we can classify the dynamics as either acting on one or several modes. We will call \textit{(single-mode) sectors} the ($q_n$, $p_n$) planes of phase space. Thus single-mode dynamics acts within a sector of phase space, while multi-mode dynamics acts across sectors.

Recall that in the matrix representations used in this paper \eqref{XhatDef}, canonical pairs of observables are adjacent. Thus, dividing $A$ and $C$ into $2$ by $2$ blocks, we can identify dynamics as being either single-mode or multi-mode depending on whether they are block on- or off-diagonal respectively. Since $\bm{b}$ can be decomposed into a sum of terms each acting within a sector without mixing, it can be identified as entirely single-mode. 

\subsection{Construction of the Partition}
Now that we have clarified our four dichotomies we can explicitly partition the terms in the Gaussian master equations \eqref{GeneralDiffXUp} and \eqref{GeneralDiffVUp} along these lines. 

For organizational convenience we define the following labels: symplectic passive (SP), symplectic active (SA), unsymplectic active (UA), and unsymplectic passive (UP). The distinction between state-dependent and state-independent dynamics ($A$ versus $\bm{b}$ or $C$) is obvious and is thus left unlabeled. The distinction between single-mode and multi-mode dynamics is also left unlabeled.

As discussed above, partitioning the dynamics into its state-dependent and state-independent parts ($A$ versus $\bm{b}$ or $C$) is trivial. Within this, we will now divide the state-dependent dynamics ($A$) into various parts. 

As discussed above, we can identify the single- and multi-mode parts of $A$ by dividing $A$ into $2$ by $2$ blocks. The single-mode part of $A$ is then all the block on-diagonal entries, whereas the multi-mode part of $A$ are the block off-diagonal entries.

To make this distinction clearer, it is helpful to introduce a basis for the real $2$ by $2$ matrices,
\bel{2by2basis}
\openone_2
=\begin{pmatrix}
1 & 0 \\
0 & 1 \\
\end{pmatrix},
\,
\omega=\begin{pmatrix}
0 & 1 \\
-1 & 0 \\
\end{pmatrix},
\, 
X=\begin{pmatrix}
0 & 1 \\
1 & 0 \\
\end{pmatrix},
\,
Z=\begin{pmatrix}
1 & 0 \\
0 & -1 \\
\end{pmatrix}
\ee
and expand $A$ over this basis in the second tensor factor as, 
\bel{AExpand2}
A=A_I \otimes\openone_2
+A_w \otimes\omega
+A_x \otimes X
+A_z \otimes Z.
\ee
Note that each $A_\mu$ are $N\times N$ matrices for $\mu\in\{I,w,x,z\}$ and that the rows and columns of each $A_\mu$ address individual modes. Thus we can identify the single- and multi-mode parts of $A_\mu$ as its diagonal and off-diagonal entries, respectively. 

Explicitly, defining $A_\mu^D$ as the matrix with the same diagonal entries as $A_\mu$ and all other entries equal to zero we have
\begin{align}
\text{Single-mode part of } A&=A_I^D \otimes\openone_2
+A_w^D \otimes\omega\\
\nonumber
&+A_x^D \otimes X
+A_z^D \otimes Z.
\end{align}
The multi-mode part of $A$ is given by the difference of $A$ and its single mode part.

In the previous subsection, we partitioned $A$ into its symplectic and unsymplectic parts, $A_\textsc{s}$ and $A_\textsc{u}$, in equations \eqref{AsDef} and \eqref{AuDef} by isolating its symmetric and antisymmetric parts. Likewise we partitioned $A$ into its active and passive parts, $A_\textsc{a}$ and $A_\textsc{p}$, in equations \eqref{AaDef} and \eqref{ApDef} by isolating the symmetric and antisymmetric parts of $\Omega A$. A priori there is no reason to expect that this second symmetrization (of $\Omega A$) respects the first one (of $\Omega$). However if we compute the passive and symplectic part of $A$ by composing these two symmetrizations as 
\begin{align}
A_\textsc{sp}
\nonumber
&=\Omega^{-1}(\Omega A_\textsc{s}+(\Omega A_\textsc{s})^\intercal)/2\\
&=(A+A^\intercal
-\Omega^{-1} A^\intercal \Omega^\intercal
-\Omega^{-1} A \Omega^{\intercal})/4.
\end{align}
we do in fact find that $A_\textsc{sp}$ is still symmetric (and therefore still symplectic). Note that this only follows because $\Omega$ has $\Omega^{-1}=\Omega^\intercal$.

One can check that the same holds for the other three pairs of symmetries and even reversing their order of application. Thus we can define
\begin{align}
A_\textsc{sp}
&\coloneqq\Omega^{-1}(\Omega A_\textsc{s}+(\Omega A_\textsc{s})^\intercal)/2
=(A_\textsc{p}+A_\textsc{p}^\intercal)/2\\
A_\textsc{sa}
&\coloneqq\Omega^{-1}(\Omega A_\textsc{s}-(\Omega A_\textsc{s})^\intercal)/2
=(A_\textsc{a}+A_\textsc{a}^\intercal)/2\\
A_\textsc{ua}
&\coloneqq\Omega^{-1}(\Omega A_\textsc{u}-(\Omega A_\textsc{u})^\intercal)/2
=(A_\textsc{a}-A_\textsc{a}^\intercal)/2\\
A_\textsc{up}
&\coloneqq\Omega^{-1}(\Omega A_\textsc{u}+(\Omega A_\textsc{u})^\intercal)/2
=(A_\textsc{p}-A_\textsc{p}^\intercal)/2.
\end{align}

In order to find a more convenient form for these expressions we first note that the symplectic form can be written as \mbox{$\Omega=\openone_N\otimes\omega$}. That is, it acts trivially on the first tensor factor and by $\omega$ in the second factor. 

As we did in \eqref{AExpand2} we can expand $A$ over the 2 by 2 basis \eqref{2by2basis} in its second tensor factor as, 
\be
A=A_I \otimes\openone_2
+A_w \otimes\omega
+A_x \otimes X
+A_z \otimes Z
\ee
where each $A_\mu$ are $N\times N$ matrices for $\mu\in\{I,w,x,z\}$.

Using this expansion we now isolate each part of $A$ described above. For instance to find the symplectic active part of $A$ we first isolate the symplectic part of $A$ \eqref{AsDef}, that is its symmetric part. Defining 
\be
\text{sym}(A)=(A+A^\intercal)/2 \ \ \text{and}\ \ \text{anti}(A)=(A-A^\intercal)/2,
\ee 
we can compute
\begin{align}
A_\textsc{s}
&=\text{sym}(A)\\
&\nonumber
=\text{sym}(A_I \otimes\openone_2
+A_w \otimes\omega
+A_x \otimes X
+A_z \otimes Z)\\
&\nonumber
=\text{sym}(A_I \otimes\openone_2)
+\text{sym}(A_w \otimes\omega)\\
&\nonumber
+\text{sym}(A_x \otimes X)
+\text{sym}(A_z \otimes Z).
\end{align}
To further symplify this expression we note the following identities for symmetrizing tensor products: If $Y$ is symmetric,
\begin{align}\label{YsymId}
\text{sym}(A_\mu \otimes Y)
&=\text{sym}(A_\mu)\otimes Y.
\end{align}
Similarly if $Y$ is antisymmetric,
\begin{align}\label{YantiId}
\text{sym}(A_\mu \otimes Y)
&=\text{anti}(A_\mu)\otimes Y.
\end{align}
Using these identities and the symmetries of our basis elements we find
\begin{align}
A_\textsc{s}
&=\text{sym}(A_I \otimes\openone_2)
+\text{sym}(A_w \otimes\omega)\\
&\nonumber
+\text{sym}(A_x \otimes X)
+\text{sym}(A_z \otimes Z)\\
&\nonumber
=\text{sym}(A_I)\otimes\openone_2
+\text{anti}(A_w)\otimes\omega\\
&\nonumber
+\text{sym}(A_x) \otimes X
+\text{sym}(A_z) \otimes Z.
\end{align}

We can now isolate the active part of $A$ within this by multiplying by $\Omega$ symmetrizing and multipling by $\Omega^{-1}$ as 
\be
A_\textsc{sa}
=\Omega^{-1}\text{sym}(\Omega A_\textsc{s}).
\ee
Note that multiplying by $\Omega=\openone_N\otimes\omega$ has no effect on the coefficient matrices, $A_\mu$, and has the effect of permuting the 2 by 2 basis elements as
\be
\Omega: \quad I\to\omega, \ \ \ \omega\to-I, \ \  \ X\to Z, \ \ \ Z\to -X.
\ee
Using this we can calculate
\begin{align}
\Omega A_\textsc{s}
&=\text{sym}(A_I)\otimes\omega
-\text{anti}(A_w)\otimes\openone_2\\
&\nonumber
+\text{sym}(A_x) \otimes Z
-\text{sym}(A_z) \otimes X
\end{align}
Using \eqref{YsymId} and \eqref{YantiId} we next calculate,
\begin{align}
\text{sym}(\Omega A_\textsc{s})
&\nonumber
=\text{anti}(\text{sym}(A_I))\otimes\omega
-\text{sym}(\text{anti}(A_w))\otimes\openone_2\\
&\nonumber
+\text{sym}(\text{sym}(A_x)) \otimes Z
-\text{sym}(\text{sym}(A_z)) \otimes X
\end{align}
Finally we multiply by $\Omega^{-1}$ reversing the previous permutation. This yields
\begin{align}
A_\textsc{sa}
&=\Omega^{-1}\text{sym}(\Omega A_\textsc{s})\\
&\nonumber
=\text{anti}(\text{sym}(A_I))\otimes\openone_2
+\text{sym}(\text{anti}(A_w))\otimes\omega\\
&\nonumber
+\text{sym}(\text{sym}(A_x)) \otimes X
+\text{sym}(\text{sym}(A_z)) \otimes Z.
\end{align}
Using the fact that $\text{sym}$ and $\text{anti}$ are orthogonal projectors we 
arrive at the final expression
\begin{align}
A_\textsc{sa}
&=\text{sym}(A_x) \otimes X
+\text{sym}(A_z) \otimes Z.
\end{align}
Similar calculations for the other parts of $A$ yield
\begin{align}
\label{ASPDef}
A_\textsc{sp}
&=A_{I,\text{sym}}\otimes\openone_2
+A_{w,\text{anti}}\otimes\omega\\
\label{ASADef}
A_\textsc{sa}
&=A_{x,\text{sym}}\otimes X
+A_{z,\text{sym}}\otimes Z\\
\label{AUADef}
A_\textsc{ua}
&=A_{I,\text{anti}}\otimes\openone_2
+A_{w,\text{sym}}\otimes\omega\\
\label{AUPDef}
A_\textsc{up}
&=A_{x,\text{anti}}\otimes X
+A_{z,\text{anti}}\otimes Z
\end{align}
where $A_{\mu,\text{sym}}=(A_\mu+A_\mu^\intercal)/2$ and $A_{\mu,\text{anti}}=(A_\mu-A_\mu^\intercal)/2$. That is, $A$ breaks up into combinations of $N\times N$ symmetric and antisymmetric matrices tensored with elements of the basis \eqref{2by2basis}.

Expressed in this form we can easily identify the single- and multi-mode parts of each of these terms by recalling that the block diagonal entries of these matrices (and thus the diagonal entries of their first tensor factors) correspond to single-mode dynamics. For instance the single-mode part of $A_\textsc{sp}$ is given by,
\begin{align}
\nonumber
\text{Single-mode part of } A_\textsc{sp}
&=A_{I,\text{sym}}^D\otimes\openone_2
+A_{w,\text{anti}}^D\otimes\omega.
\end{align}
Recall that $A_\mu^D$ is the matrix with the same diagonal entries as $A_\mu$ and all other entries equal to zero. Note that $A_{w,\text{anti}}$ is antisymmetric such that $A_{w,\text{anti}}^D=0$ and 
\begin{align}
\text{Single-mode part of } A_\textsc{sp}
&=A_{I,\text{sym}}^D\otimes\openone_2.
\end{align}
The multi-mode part of $A_\textsc{sp}$ is given by the difference of $A_\textsc{sp}$ and its single mode part. Similarly we can isolate the single-mode and multi-mode parts of $A_\textsc{sa}$, $A_\textsc{up}$, and $A_\textsc{ua}$.

Note that since both of the first tensor factors in $A_\textsc{up}$ are antisymmetric they neither have diagonal entries. Thus the single-mode part of $A_\textsc{up}$ vanishes,
\begin{align}\label{AupSingle0}
\text{Single-mode part of } A_\textsc{up}
=A_{x,\text{anti}}^D\otimes X
+A_{z,\text{anti}}^D\otimes Z
=0.
\end{align}
Thus it is not possible to have single-mode dynamics which is unsymplectic, passive, and state-dependent. The other three parts of $A$ can generically have both single- and multi-mode parts.

Now that we have fully partitioned the state-dependent part of the dynamics ($A$), we turn our attention to the state-independent part of the dynamics ($\bm{b}$ and $C$). 

As discussed throughout Sec. \ref{Criteria} the contribution to the dynamics coming from $\bm{b}$ is entirely characterized as symplectic, active, state-independent, and single-mode. There is no further division to be done on $\bm{b}$. 

The $C$ term on the other hand is always unsymplectic and state-independent, but can be either single- or multi-mode as well as either active or passive. We will now complete the partition of $C$ along these lines. 

Just as with $A$, in order to identify the single- and multi-mode parts of $C$, we expand it over the basis \eqref{2by2basis} in its second tensor factor as,
\be
C=C_{I}\otimes\openone_2
+C_{w}\otimes\omega
+C_{x}\otimes X
+C_{z}\otimes Z.
\ee
Note that each $C_\mu$ are $N$ by $N$ matrices for $\mu\in\{1,w,x,z\}$. Since $C$ is always symmetric, $C_{I}$, $C_{w}$, $C_{x}$ and $C_{z}$ obey some symmetry relationships as well. Specifically, since $\openone_2$, $X$ and $Z$ are symmetric so must be $C_{I}$, $C_{x}$, and $C_{z}$. Furthermore, since $\omega$ is antisymmetric, $C_{w}$ must be antisymmetric as well.

Using this we can identify the single- and multi-mode parts of each $C_\mu$ as its diagonal and off-diagonal entries, respectively. Thus  we can identify
\begin{align}
\nonumber
&\text{Single-mode part of } C\\
\nonumber
&=C_{I}^D\otimes\openone_2
+C_{w}^D\otimes\omega
+C_{x}^D\otimes X
+C_{z}^D\otimes Z\\
&=C_{I}^D\otimes\openone_2
+C_{x}^D\otimes X
+C_{z}^D\otimes Z
\end{align}
where $C_{\mu}^D$ is a matrix with the same diagonal entries as $C_{\mu}$ and all other entries equal to zero. Note that since $C_{w}$ is antisymmetric $C_{w}^D=0$. The multi-mode part of $C$ is the difference between $C$ and its single-mode part.

Finally, we can divide $C$ into its active and passive parts. As we saw from \eqref{Activity}, $C$ contributes to the creation/annihilation of excitations through its trace. The multi-mode part of $C$ is block off-diagonal and therefore does not contribute to the trace and is thus passive. This implies that the active part of $C$ is entirely single-mode. Within the single-mode part of $C$, we can identify $C_{x}$ and $C_{z}$ as being passive as well, since $X$ and $Z$ are traceless. Thus we are lead to the conclusion that the only active part of $C$ is the diagonal entries of $C_{I}$. Explicitly,
\begin{align}
\label{CUADef}
C_\textsc{ua}
&=C_{I}^D\otimes\openone_2\\
\label{CUPDef}
C_\textsc{up}&=C-C_\textsc{ua}.
\end{align}

The results of the partition implemented above are summarized in Table \ref{SUAPTable}.
\begin{table}[h]
\begin{tabular}{|r|c|c|c|c|} 
\hline
& \multicolumn{2}{c|}{\bf Active} & \multicolumn{2}{c|}{\bf Passive}\\ 
\hline
{\bf Symplectic} 
& \ $A_\textsc{sa}$  ${}^{(s/m)}$
& \ $\bm{b}$ \ \ \, ${}^{(s/\,\ )}$
& \ $A_\textsc{sp}$  ${}^{(s/m)}$
& \quad\quad\quad \\
\hline
{ \bf Unsymplectic} 
& \ $A_\textsc{ua}$  ${}^{(s/m)}$
& \ $C_\textsc{ua}$  ${}^{(s/\,\ )}$
& \ $A_\textsc{up}$  ${}^{(\,\ /m)}$
& \ $C_\textsc{up}$  ${}^{(s/m)}$ \\ 
\hline
\quad  
& \quad \bf S.D. \quad
& \quad \bf S.I. \quad
& \quad \bf S.D. \quad
& \quad \bf S.I. \quad\\ 
\hline
\end{tabular}
\caption{The result of the partition described above. Note that the horizontal division within each cell indicates state dependence (S.D.) or independence (S.I.). The parenthetical note indicates if the dynamics can be either single-mode (s) or multi-mode (m). Note that there is no symplectic, passive, and state-independent dynamics, either single- or multi-mode.}
\label{SUAPTable}
\end{table}

\subsection{Analysis of Partition}
\subsubsection{``Missing'' Dynamics}
One may have expected that dividing open Gaussian dynamics along the four dichotomies described in Section \ref{Criteria} would yield 16 different types of dynamics. However, from Table \ref{SUAPTable} we see that there are only 11 types of open Gaussian dynamics. The four dichotomies we have used in constructing our partition are not strictly independent.

For example, since $C$ is always unsymplectic, dynamics that are symplectic and state-independent must come from $\bm{b}$ but as discussed above $\bm{b}$ is necessarily active and single-mode. Thus we have the following implication:
\be
\nonumber
\text{Symplectic, state-independent}
\Rightarrow
\text{active, single-mode}.
\ee
This implies in turn that the following three types of dynamics cannot exist:
\begin{align}
    &\nonumber
    \text{Symplectic, state-independent, passive, single-mode;}\\
    &\nonumber
    \text{Symplectic, state-independent, active, multi-mode;}\\
    &\nonumber
    \text{Symplectic, state-independent, passive, multi-mode.}
\end{align}
Similarly, any dynamics which are unsymplectic, state-independent, and active must come from $C_\textsc{UA}$. However, from \eqref{CUADef} we can see that $C_\textsc{UA}$ is diagonal and therefore single-mode. Thus we have the implication
\be
\nonumber
\text{unsymplectic, state-independent, active}
\Rightarrow
\text{single-mode.}
\ee
Therefore, there is no dynamics which is, unsymplectic, state-independent, active, and multi-mode.

Finally, as noted in equation \eqref{AupSingle0}, $A_\text{up}$ is necessarily multi-mode. Thus we have
\be
\nonumber
\text{unsymplectic, passive, state-dependent}
\Rightarrow
\text{multi-mode}
\ee
such that there is no dynamics which is, unsymplectic, passive, state-dependent, and single-mode.

The remaining 11 types of dynamics are logically independent. They are listed and given names in Table \ref{Table2} according to their effect on an arbitrary Gaussian state (see Section \ref{Naming}). Roughly, the $\bm{b}$ term implements displacement, the $C$ term adds noise to the dynamics, and $A$ terms have several effects including rotations, squeezings, and amplifications.

\begin{table*}
\begin{tabular}{||cccc||c||}
\hline 
\quad Single-mode? \quad & \quad Symplectic? \quad  &  \quad Passive? \quad  &  \quad State-Dependent? \quad  &            \\
 \quad (else Multi-mode) \quad  &  \quad (else Unsymplectic) \quad  &  \quad (else Active) \quad  &  \quad (else Independent) \quad  &       Name \\
\hline Yes &        Yes &        Yes &        Yes & Single-mode Rotation \\
\hline Yes &        Yes &        Yes &         No &        Not Possible \\
\hline Yes &        Yes &         No &        Yes & Single-mode Squeezing \\
\hline Yes &        Yes &         No &         No & Displacement \\
\hline Yes &         No &        Yes &        Yes &        Not Possible \\
\hline Yes &         No &        Yes &         No & Single-mode Squeezed Noise \\
\hline Yes &         No &         No &        Yes & Amplification/Relaxation \\
\hline Yes &         No &         No &         No & Free Thermal Noise \\
 \hline No &        Yes &        Yes &        Yes & Multi-mode Rotation \\
 \hline No &        Yes &        Yes &         No &        Not Possible \\
 \hline No &        Yes &         No &        Yes & Multi-mode Squeezing \\
 \hline No &        Yes &         No &         No &        Not Possible \\
 \hline No &         No &        Yes &        Yes & Multi-mode Counter-Rotation \\
 \hline No &         No &        Yes &         No & Multi-mode Squeezed Noise \\
 \hline No &         No &         No &        Yes & Multi-mode Counter-Squeezing \\
 \hline No &         No &         No &         No &        Not Possible \\
\hline
\end{tabular}
\caption{The result of our division. Eleven of the possible sixteen types of dynamics are logically possible. These are named in this table. See Sec. \ref{Naming} for phase space plots of each type of dynamics.
}\label{Table2}
\end{table*}

\subsubsection{Complete Positivity}\label{GaussianCP}
While the 11 remaining types of dynamics are logically independent in general, enforcing that the Gaussian master equations \eqref{GeneralDiffXUp} and \eqref{GeneralDiffVUp} are completely positive induces new relationships between different types of dynamics. In particular in order to be completely positive the total dynamics must obey the complete positivity conditions \eqref{FiniteCPCond},
\bel{DiffCPCond2}
C\geq\ii \Omega (A-A^\intercal)\Omega
\ee
(see \eqref{DiffCPCond} and recall that the notation $P\geq Q$ here means that $P-Q$ is positive semi-definite). Many of the types of dynamics arising from the above partition do not satisfy this inequality in isolation.

As we will discuss below, \eqref{DiffCPCond2} implies two useful facts: $C\geq0$, and if $A\neq A^\intercal$ then $C\neq0$ and $\text{Tr}(C)>0$. The first fact (that $C\geq0$ for completely positive dynamics) implies that $C$ can only increase the uncertainty of the state. The second fact means that unsymplectic dynamics require noise ($C\neq0$) to be completely positive and moreover this noise will be active.

The proof of the first statement above can be derived from the same argument following \eqref{SigmaPosCond}. Namely, by the postivity of conjugation
\be
C\geq\ii \Omega (A-A^\intercal)\Omega
\Rightarrow
C\geq-\ii \Omega (A-A^\intercal)\Omega
\ee
since $C$ and $\Omega (A-A^\intercal)\Omega$ are real-valued matrices. Therefore by the convexity of positive semi-definite matrices, $C\geq0$.

To prove the second implication, first suppose that the dynamics is unsymplectic, that is $A\neq A^\intercal$ or $C\neq0$. As we will show below, in either case we have $C\neq0$ and therefore since $C\geq0$ we have \mbox{$\text{Tr}(C)>0$}, such that the dynamics is active.

First show that note $A\neq A^\intercal$ implies $C\neq0$ we first note that $\ii \, \Omega (A-A^\intercal)\Omega$ is imaginary valued and antisymmetric. Thus its eigenvalues are real and come in pairs $\pm\lambda$. Assuming $A\neq A^\intercal$, we can thus see that $\ii \, \Omega (A-A^\intercal)\Omega\neq0$ has both positive and negative eigenvalues and is therefore neither positive nor negative semi-definite. On the other hand, if the dynamics is completely postive and $C=0$ then \eqref{DiffCPCond2} would directly yield $0\geq \ii \, \Omega (A-A^\intercal)\Omega$, such that $\ii \, \Omega (A-A^\intercal)\Omega$ is negative semi-definite, contradicting our earlier conclusion. Thus if $A\neq A^\intercal$ then $C\neq0$.

Thus in order to be completely positive unsymplectic dynamics requires active noise, i.e., $C_\textsc{ua}\neq0$. From \ref{SUAPTable} we can see that such dynamics is necessarily single-mode. From Table \ref{Table2} we can identify this type of dynamics (unsymplectic, state-independent, active and single-mode) as free thermal noise. Thus any unsymplectic dynamics require the presence of thermal noise to be completely positive.

Note that since thermal noise is active, this means that any completely positive unsymplectic dynamics will be active in total as well.

From this we can make an interesting observation regarding the relationship between the flows of information and the flows of energy in Gaussian Quantum Mechanics. Recall that the generation of entanglement between the system and its environment necessitates non-unitary dynamics on the system. In the Gaussian regime such dynamics is completely positive and unsymplectic. As we have just seen this type of Gaussian dynamics must be active. This means that it is not possible to guarantee that a Gaussian quantum channel that generates entanglement will always have zero energy flow. In general we should expect that Gaussian quantum channels that allow for quantum information to \textit{leak} to the environment (understood as system-environment entanglement generation) will also allow for a system-environment energy flow.

We conjecture that this result could be due to our assumption that each of the modes has a finite energy scale. The inclusion of zero-modes would add many distinguishable states to the system which are all energetically equivalent. Entangling these states should not require energy flows.

\subsubsection{Conservation of Phase Space Volume}
We can also analyze dynamics based on its effect on the volume of phase space. To this end we define the volume form on phase space as
\begin{align}\label{VolDef}
V&\coloneqq\d q_1\wedge \d p_1\wedge\dots\wedge\d q_N\wedge\d p_N\\
\nonumber
&=\frac{1}{N!}\Omega\wedge\dots\wedge\Omega
=\frac{1}{N!}\Omega^{\wedge N}
\end{align}
where $\Omega$ is the symplectic form, i.e., the two form 
\be
\Omega=\sum_{n=1}^N \d q_n\wedge\d p_n,
\ee
where $\wedge$ denotes the wedge product and $\d q_n$ and $\d p_n$ are differential forms.

Thus any transformation which preserves the symplectic form will also preserve the volume of phase space. As we will see in this subsection, the converse is not strictly true. There are transformations which preserve the volume of phase space, but do not preserve the symplectic form. If, however, we demand that the transformations are completely positive, the converse does indeed hold.

The volume of a Gaussian state's ellipsoid in phase space is given, through \eqref{VolDef}, by its determinant. Specifically
\be
\text{Vol}(\bm{X},\sigma)
\coloneqq
\sqrt{\text{det}(\sigma)}.
\ee
Assuming $\sigma$ is invertible (i.e., the state is not infinitely squeezed) and using Jacobi's formula and the linear algebra identity,
\be
M \text{adj}(M)=\text{det}(M)\openone, \ee 
and \eqref{GeneralDiffVUp} we can calculate rate of change of this volume as
\begin{align}
\label{VolComputation}
&\frac{\d}{\d t}\text{Vol}(\bm{X}(t),\sigma(t))\\
\nonumber
&=\frac{1}{2}\text{det}\big(\sigma(t)\big)^{-1/2} \ \frac{\d}{\d t}\text{det}\big(\sigma(t)\big)\\
\nonumber
&=\frac{1}{2}\text{det}\big(\sigma(t)\big)^{1/2} \ \text{Tr}\big(\sigma^{-1}(t)\frac{\d}{\d t}\sigma(t)\big)\\
\nonumber
&=\frac{1}{2}\text{det}\big(\sigma(t)\big)^{1/2} \ \text{Tr}\Big(\sigma^{-1}(t)\big(\Omega A \sigma(t)+\sigma(t)(\Omega A)^\intercal+C\big)\Big)\\
\nonumber
&=\frac{1}{2}\text{det}\big(\sigma(t)\big)^{1/2} \ \text{Tr}\big(\Omega A+(\Omega A)^\intercal+\sigma^{-1}(t) \, C\big)\\
\nonumber
&=\frac{1}{2}\text{det}\big(\sigma(t)\big)^{1/2} \ \text{Tr}\big(2 \, \Omega A+\sigma^{-1}(t) \, C\big).
\end{align}
In order for this volume to be conserved for all states, we would need $\Omega A$ to be traceless and $C=0$. Thus all types of noise do not conserve the volume of phase space, as well as any parts of $A$ contributing to the trace of $\Omega A$.

We now determine which parts of $A$ contribute to the non-conservation of phase space volume. As mentioned above, the volume form, \eqref{VolDef}, can be built out of the symplectic form. Thus the symplectic part of $A$, $A_\textsc{s}$, conserves the volume of phase space. Thus any non-conservation of phase space volume coming from $A$ will necessarily be unsymplectic.

Furthermore the passive part of $A$, $A_\textsc{p}$, conserves the volume of phase space since by definition $\Omega A_\textsc{p}$ is antisymmetric and therefore traceless. Thus any non-conservation of phase space volume coming from $A$ will necessarily be active.

Finally, we note that multi-mode parts of $A$ (the block off-diagonal parts of $A$) remain block off-diagonal when multiplied by $\Omega$, since $\Omega$ itself is block diagonal. Since block off-diagonal matrices are traceless the multi-mode parts of $A$ cannot contribute to the trace of $\Omega A$, hence do not change the volume of phase space. 

In conclusion the parts of $A$ which yield non-conservation of phase space volume are unsymplectic, active, single-mode, and state-dependent. From Table \ref{Table2} we can identify this type of dynamics as amplification/relaxation.

Note that the above argument implies that all of the other parts of $A$ will conserve the volume of phase space, even including its unsymplectic parts! At first sight this may seem to run counter to the common intuition that ``unsymplectic dynamics causes non-conservation of the volume of phase space''. We can reconcile this expectation with our findings by noting that, as discussed in the previous subsection, completely positive unsymplectic dynamics always require noise. This noise will itself not conserve the volume of phase space. Thus we can qualify the previous statement by saying that ``completely positive unsymplectic dynamics causes non-conservation of the volume of phase space''. In situations where complete positivity is violated (e.g. non-Markovian dynamics \cite{nonMarkov1,nonMarkov2}) there can be unsymplectic dynamics which preserve the volume of phase space. 

\subsubsection*{Exact condition for Gaussian dynamics to purify}
We now analyze when open Gaussian dynamics in general can lead to purification in terms of the partition described above. Dynamics being able to increase the purity of at least one state is a prerequisite for the dynamics to be able to capture the process of thermalization.

Following \cite{Grimmer2017a} we say that a map can purify (reduce the entropy of the system) if there exists a state whose purity increases under the map (i.e., reduces its entropy under the map). Below we find a necessary and sufficient for Gaussian dynamics to cause purification.

The purity of a Gaussian state (see e.g., \cite{GPurity}) is given (in our notation) by
\be
\mathcal{P}=\text{Tr}(\rho^2)
=\frac{1}{\text{det}(\sigma)}.
\ee
Thus, in order for any kind of Gaussian dynamics to increase the purity of some Gaussian state, the dynamics has to decrease the determinant of its covariance matrix. Paralleling the calculation following \eqref{VolComputation} we find
\begin{align}
\frac{\d}{\d t}\text{det}\big(\sigma(t)\big)
&=\text{det}\big(\sigma(t)\big) \, \text{Tr}\big(2 \, \Omega A+\sigma^{-1}(t) \, C\big).
\end{align}
Since $\sigma\geq0$ and since we assumed $\sigma$ was nonsingular, we have $\text{det}(\sigma)>0$. Thus dynamics instantaneously increases the purity of a state with covariance $\sigma$ provided
\bel{GaussianPurificationInequality}
\text{Tr}\big(\Omega A\big)
<-\frac{1}{2}\text{Tr}\big(\sigma^{-1} \, C\big)
\leq0
\ee
where we have noted that since $C$ and $\sigma$ (and hence $\sigma^{-1}$) are positive semi-definite, we have \mbox{$\text{Tr}\big(\sigma^{-1} \, C\big)\geq0$}. Thus for dynamics to cause purification there must exist some valid state (i.e., $\sigma\geq\ii\Omega$) such that \eqref{GaussianPurificationInequality} holds.

From \eqref{GaussianPurificationInequality} we can easily see that
\bel{GaussianNandS}
\text{Tr}\big(\Omega A\big)<0
\ee
is a necessary condition for dynamics to cause purification. 

We can also see that this condition is sufficient for Gaussian dynamics to cause purification, as we will now show. Given some dynamics with $\text{Tr}\big(\Omega A\big)<0$ we can construct a state that is purified by the dynamics. Specifically we find a state with $-\frac{1}{2}\text{Tr}\big(\sigma^{-1} \, C\big)$ between $\text{Tr}\big(\Omega A\big)$ and zero  by taking $\sigma$ to be a thermal state, i.e. $\sigma=\nu \, \openone_{2N}$ with 
\bel{TboundGaussian}
\nu>\frac{\text{Tr}\big(C\big)}{-2\, \text{Tr}\big(\Omega A\big)}
\ee
thereby  satisfying \eqref{GaussianPurificationInequality}, where $\nu\geq1$. 
Hence dynamics with $\text{Tr}\big(\Omega A\big)<0$ will purify some state, specifically the sufficiently hot thermal state described above.

Thus \eqref{GaussianNandS} is the necessary and sufficient condition for Gaussian dynamics to cause purification. 

In the previous subsection we found that the only type of dynamics which has $\text{Tr}\big(\Omega A\big)\neq0$ is unsymplectic, active, single-mode, and state-dependent, i.e. amplification/relaxation. In order for dynamics to purify it must contain amplification and/or relaxation. Moreover there must be more relaxation than amplification. 

\section{Visualizing the Different Components}\label{Naming}
Having completed the partition described in the previous section, we now present each type of dynamics appearing in Table \ref{Table2} in turn. To aid intuition we make use of the visualization technique described in Sec. \ref{ReviewGQM} in which states are thought of as ellipsoids in a $2N$ dimensional phase space. It is sufficient to consider systems composed of one or two modes ($N=1,2$) in order to build illustrative examples of every type of dynamics that we have listed in Table \ref{Table2}.

\begin{figure*}
\includegraphics[width=0.3\textwidth]{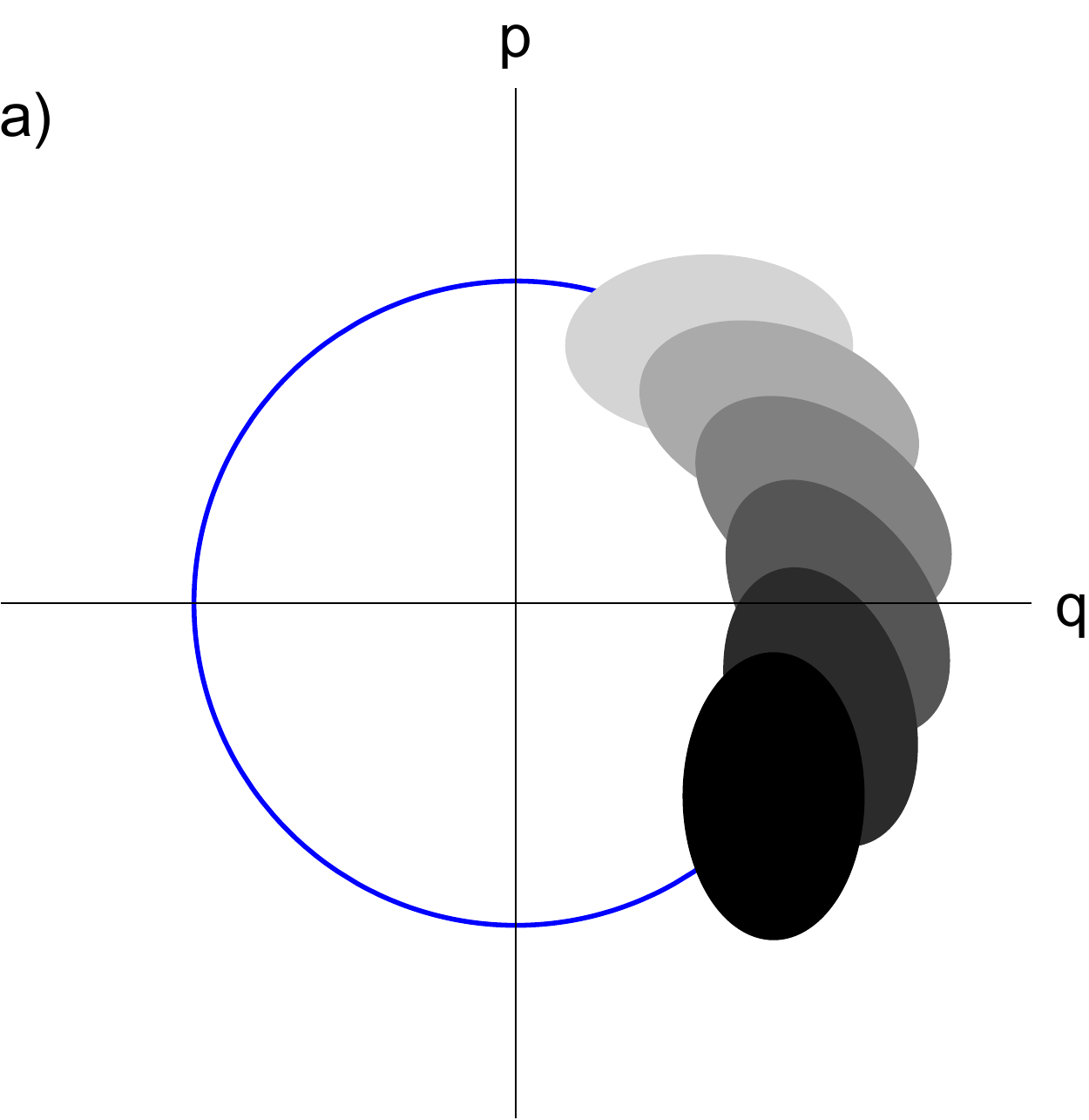}
\includegraphics[width=0.3\textwidth]{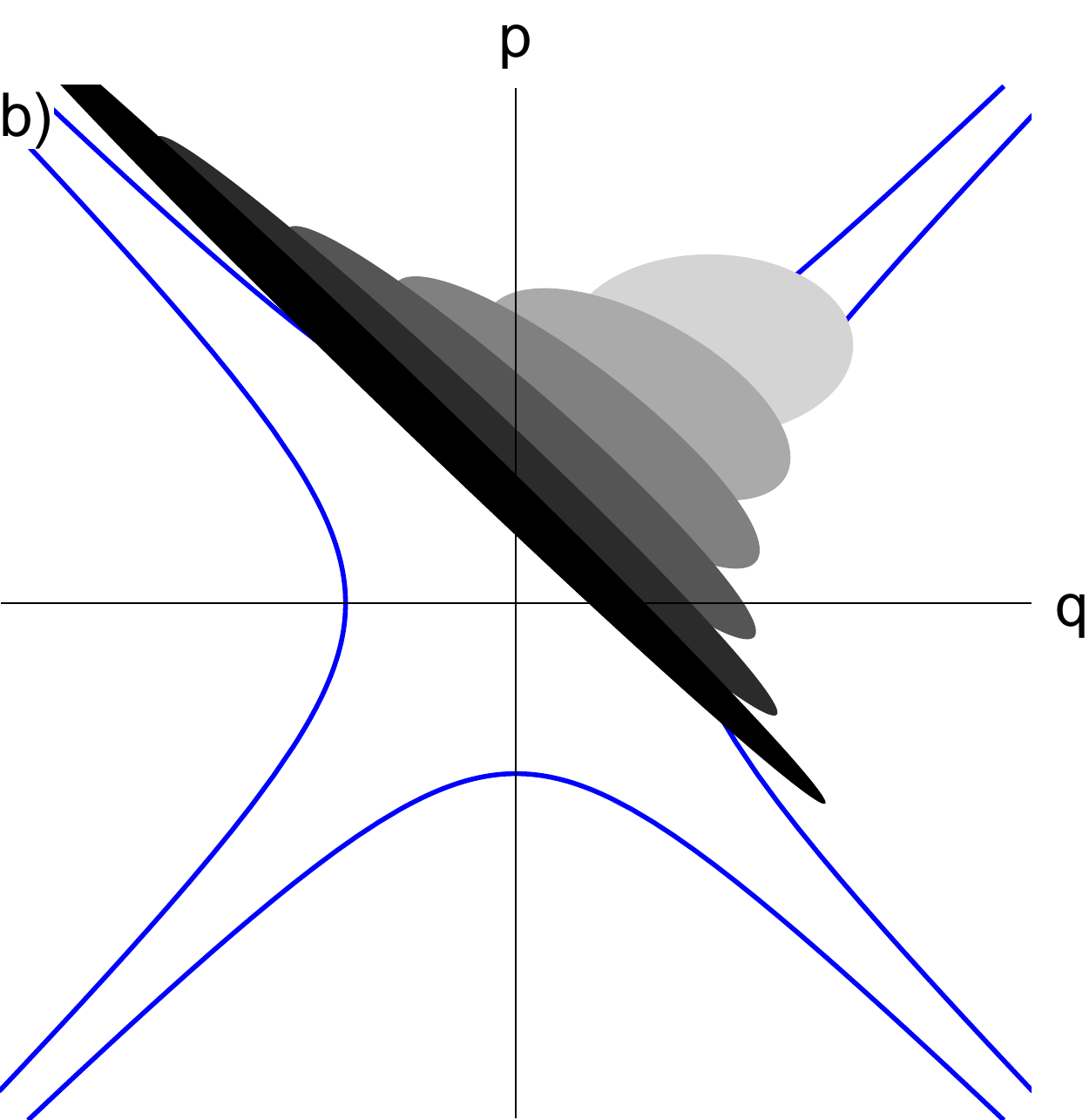}
\includegraphics[width=0.3\textwidth]{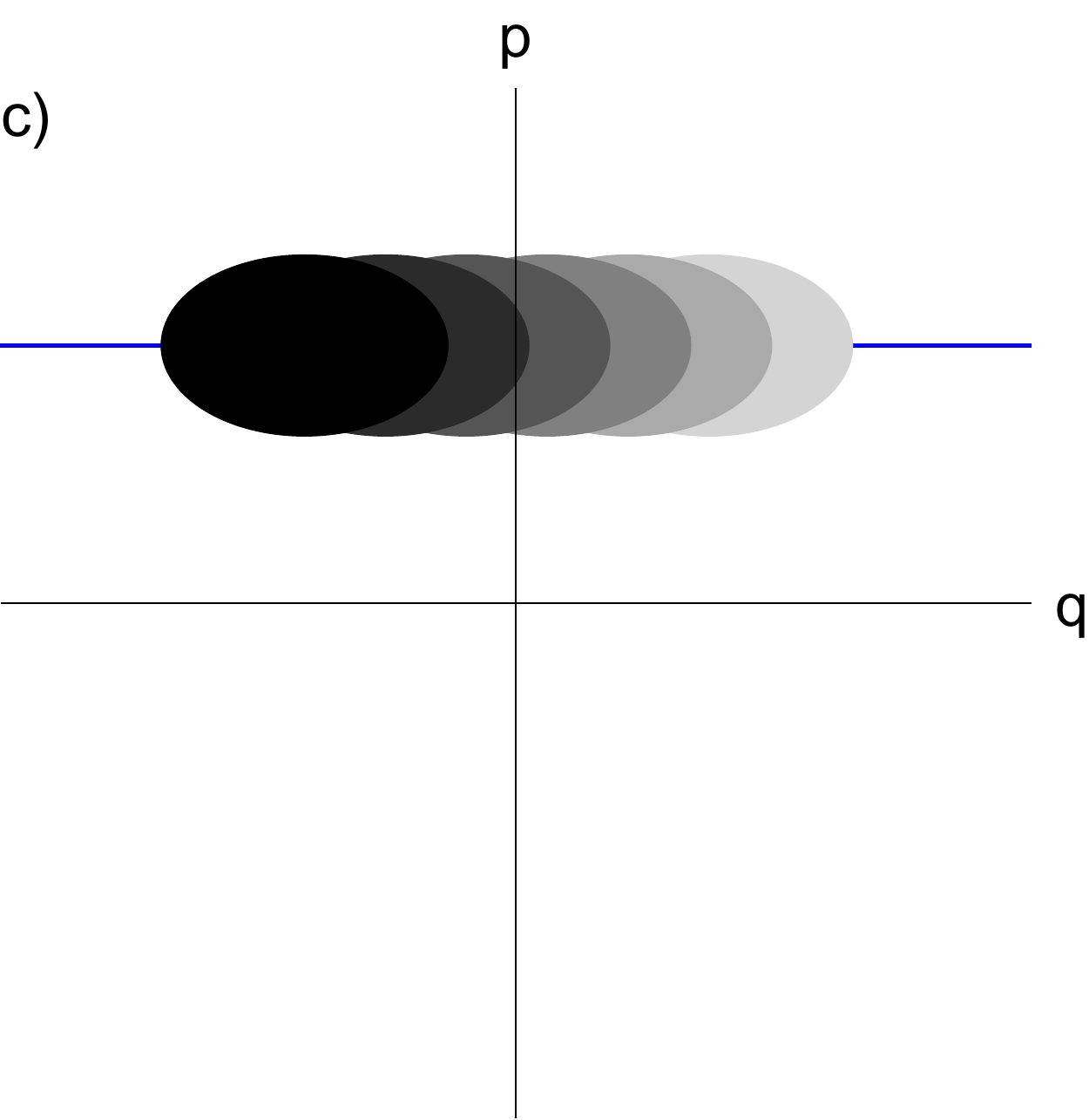}
\includegraphics[width=0.3\textwidth]{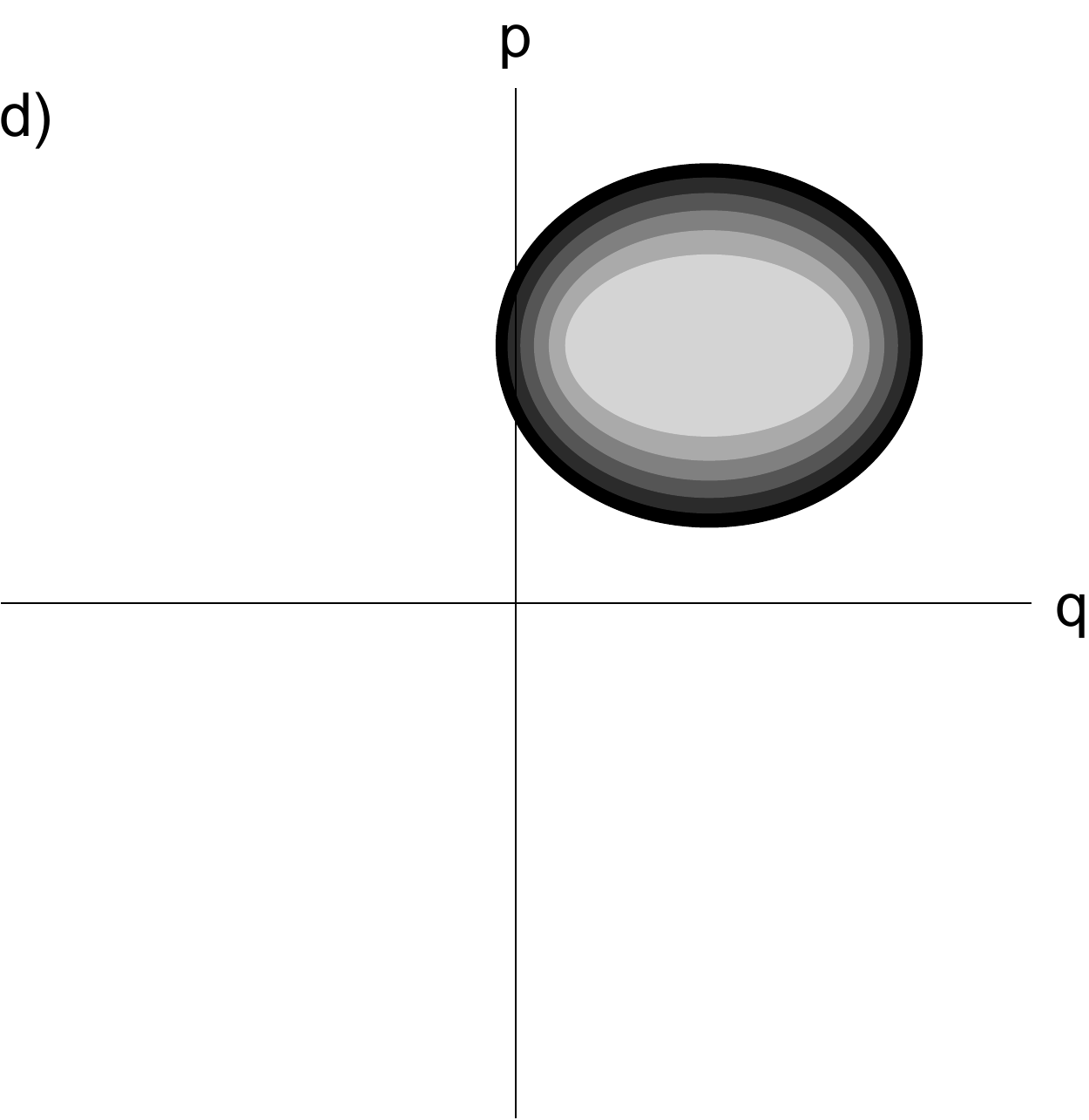}
\includegraphics[width=0.3\textwidth]{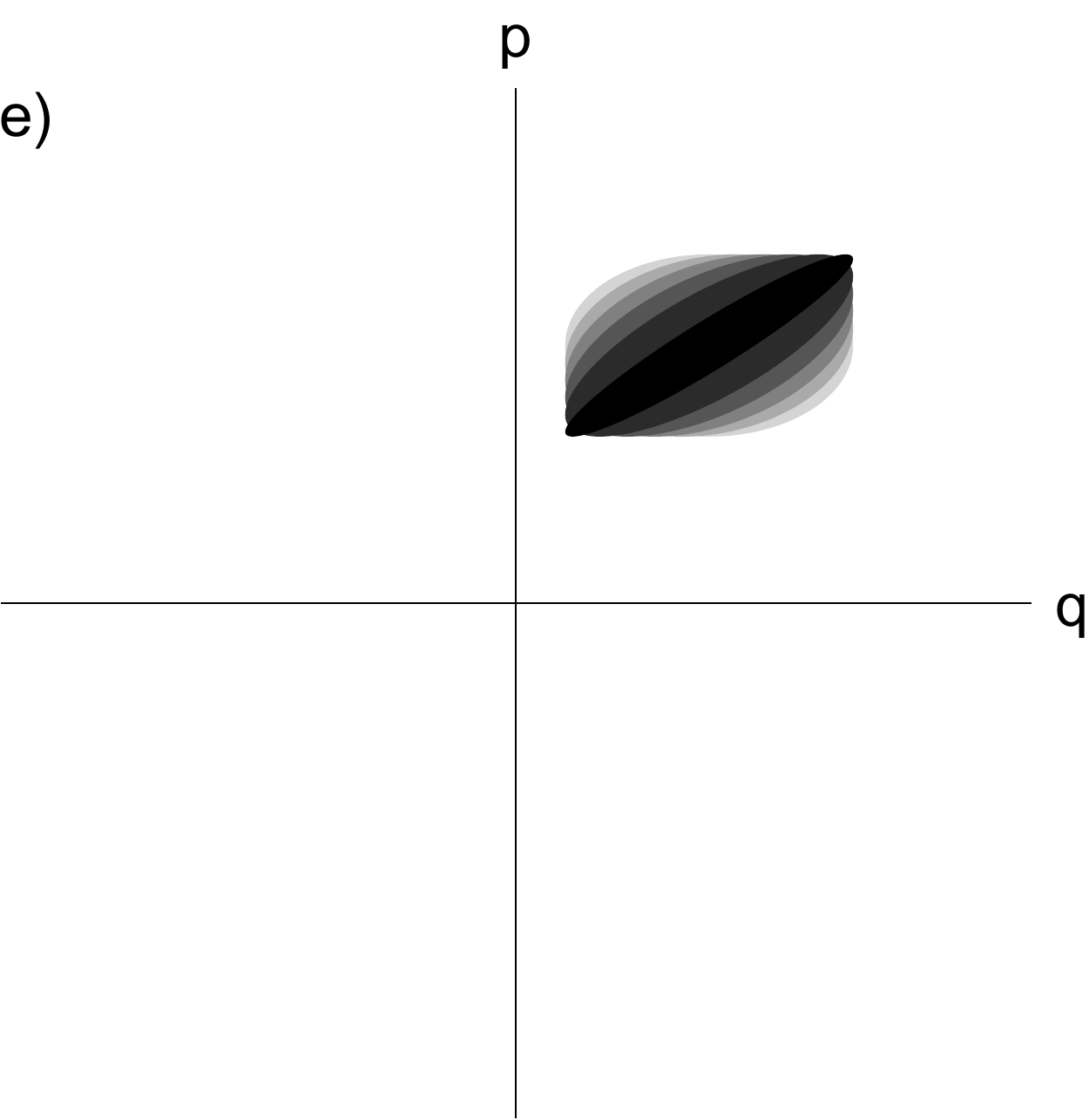}
\includegraphics[width=0.3\textwidth]{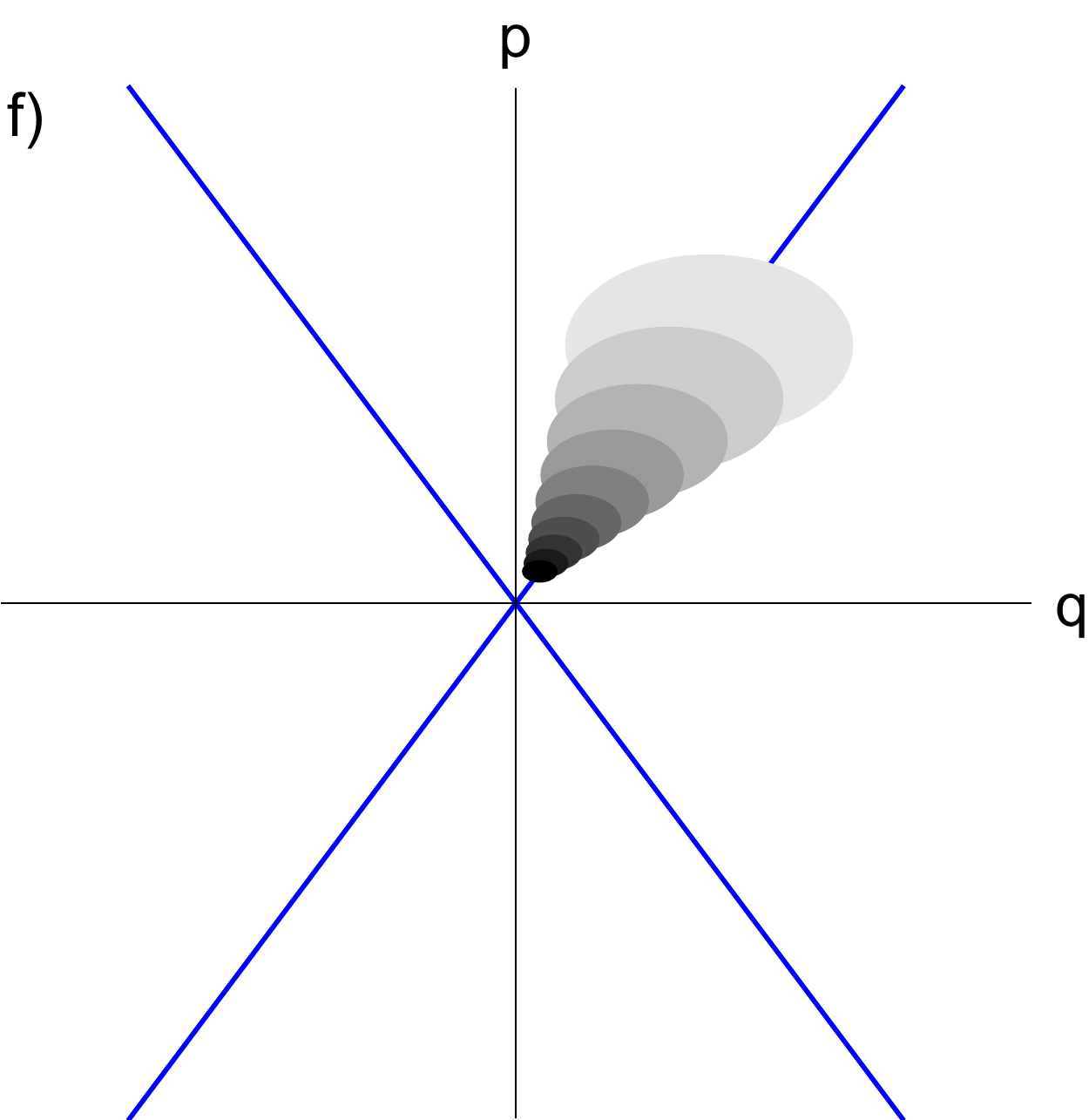}
\caption{
(Color online.) The full range of phenomena available to a single Gaussian mode. The different shades of gray indicate time evolution: lighter ellipses correspond to earlier times. The dynamics displayed in the top row are symplectic (unitary) and the bottom row are unsymplectic (nonunitary).  \textbf{a)} Single-mode rotation (section \ref{SingleModeRotation}). \textbf{b)} Single-mode squeezing (section \ref{SingleModeSqueezing}). \textbf{c)} Displacement (section \ref{Displacement}). \textbf{d)} Thermal noise (section \ref{ThermalNoise}). \textbf{e)} Squeezed Noise (section \ref{SqueezedNoise}). \textbf{f)} Amplification/Relaxation (section \ref{AmplificationRelaxation}). The path of the systems mean vector is plotted in blue. The processes e) and f) are not completely positive in isolation. They need a sufficiently large thermal noise term in order for the dynamics to be overall completely positive.}
\label{Fig:SingleModeDynamics}
\end{figure*}

\subsubsection{Single-mode Rotation}\label{SingleModeRotation}
Taking $N=1$ in equation \eqref{ASPDef} we find that the single-mode, symplectic, passive, and state-dependent dynamics is given by
\be
A_\textsc{sp}=\nu_0 \, \openone_2
=\nu_0
\begin{pmatrix}
1 & 0\\
0 & 1\\
\end{pmatrix}.
\ee
The effect this dynamics has on the ellipse in phase space is shown in Fig. \ref{Fig:SingleModeDynamics} a). Specifically, the center of the ellipse (given by the vector of means, $\bm{X}$) rotates around the origin of phase space at a rate $\nu_0$. Similarly the orientation of the ellipse (given by the covariance matrix, $\sigma$) rotates about its center at the same rate.

The Hamiltonian \eqref{HeffDef} associated with this dynamics is the free Hamiltonian of a harmonic oscillator,
\be
\hat{H}=\nu_0(\hat{q}^2+\hat{p}^2).
\ee

\subsubsection{Single-mode Squeezing}\label{SingleModeSqueezing}
Taking $N=1$ in equation \eqref{ASADef}, we find two types of single-mode, symplectic, active, and state-dependent dynamics. They are given by,
\bel{x+Squeezing}
A_\textsc{sa}=\gamma_{\times} Z
=\gamma_{\times}
\begin{pmatrix}
1 & 0\\
0 & -1
\end{pmatrix},
\quad \ \ \ 
A_\textsc{sa}=\gamma_{+} X
=\gamma_{+}
\begin{pmatrix}
0 & 1\\
1 & 0
\end{pmatrix}
\ee
for some real parameters $\gamma_{\times}$ and $\gamma_{+}$. The effect the $\times$ dynamics has on the ellipse in phase space is shown in Fig. \ref{Fig:SingleModeDynamics} b). Specifically, the center of the ellipse (given by the vector of means, $\bm{X}$) moves on a hyperbolic trajectory with asymptotes in the $\times$ directions. Similarly the ellipse's major and minor axes (given by the covariance matrix, $\sigma$) are squeezed in the $\times$ directions. Through \eqref{HeffDef}, the associated Hamiltonians are
\be
\hat{H}=\gamma_{\times}(\hat{q}^2-\hat{p}^2),
\quad\quad
\hat{H}=\gamma_{+}(\hat{q}\hat{p}+\hat{p}\hat{q}).
\ee

\subsubsection{Displacement}\label{Displacement}
Taking $N=1$ we find that $\bm{b}$, the single-mode, symplectic, active, and state-independent dynamics, is given by 
\be
\bm{b}=\begin{pmatrix}
b_q \\ b_p
\end{pmatrix}
\ee
for some real parameters $b_q$ and $b_p$. The effect this dynamics has on the ellipse in phase space is to translate the ellipse in some direction without changing its orientation. In other words it applies a state-independent displacement to the system's vector of mean $\bm{X}$ while leaving the covariance matrix, $\sigma$, unchanged. An example is shown in Fig. \ref{Fig:SingleModeDynamics} c).
The associated Hamiltonian \eqref{HeffDef} is
\be
\hat{H}=b_q \, \hat{q}+b_p \, \hat{p}.
\ee

\subsubsection{Thermal Noise}\label{ThermalNoise}
Taking $N=1$ in \eqref{CUADef} we find that the single-mode, unsymplectic, active, and state-independent dynamics is given by
\be
C_\textsc{ua}=c_\text{t} \, \openone_2
\ee
for some real parameter $c_\text{t}$. The effect the $\times$ dynamics has on the ellipse in phase space is shown in Fig. \ref{Fig:SingleModeDynamics} d). This dynamics adds isotropic noise to the system's covariance matrix, $\sigma$, while leaving its vector of means, $\bm{X}$, unchanged. 

This dynamics is unsymplectic and does not preserve the volume of phase space. From Sec. \ref{GaussianCP} we can see that in order for the dynamics to be completely positive, one needs $c_\text{t}\geq0$.

\subsubsection{Squeezed Noise}\label{SqueezedNoise}
Taking $N=1$ in \eqref{CUPDef} we find that the single-mode, unsymplectic, passive, and state-independent dynamics is given by
\be
C_\textsc{up}=c_+ \, Z
+c_\times \, X
\ee
for some real parameters $c_+$ and $c_\times$. The effect the $\times$ dynamics has on the ellipse in phase space is shown in Fig. \ref{Fig:SingleModeDynamics} e). This dynamics adds squeezed noise to the system's covariance matrix while leaving the mean vector unchanged. 

This dynamics is unsymplectic and does not preserve the volume of phase space. Recall that, as seen in Sec. \ref{GaussianCP}, such dynamics needs to be accompanied by a sufficient amount of free thermal noise (of the kind described in Sec. \ref{ThermalNoise}) in order to be completely positive.

\subsubsection{Amplification/Relaxation}\label{AmplificationRelaxation}
Taking $N=1$  in \eqref{AUADef} we find that the single-mode, unsymplectic, active, and state-dependent dynamics is given by
\be
A_\textsc{ua}=\eta \, \omega
=\eta\begin{pmatrix}
0 & 1\\
-1 & 0
\end{pmatrix}.
\ee
This dynamics causes the vector of means, $\bm{X}$, to move towards or away from the origin of phase space exponentially fast depending on whether $\eta$ is positive or negative respectively. Similarly the size of the ellipse (given by the eigenvalues of the covariance matrix, $\sigma$) expands or contracts exponentially quickly. The effect that this dynamics (with $\eta>0$) has on the ellipse in phase space is shown in Fig. \ref{Fig:SingleModeDynamics} f).

The effect of $A_\textsc{ua}$ on the state of the system is unsymplectic and does not preserve the volume of phase space. Whether $\eta$ is positive of negative, the evolution generated by $A_\textsc{ua}$ is not completely positive on its own. As seen in Sec. \ref{GaussianCP}, in order for the dynamics generated by $A_\textsc{ua}$ to be completely positive it must be accompanied by a sufficiently large free thermal noise term (of the kind seen in Sec. \ref{ThermalNoise}).

\subsubsection{Multi-mode Rotation}\label{MultiModeRotation}
\begin{figure*}
\includegraphics[width=0.48\textwidth]{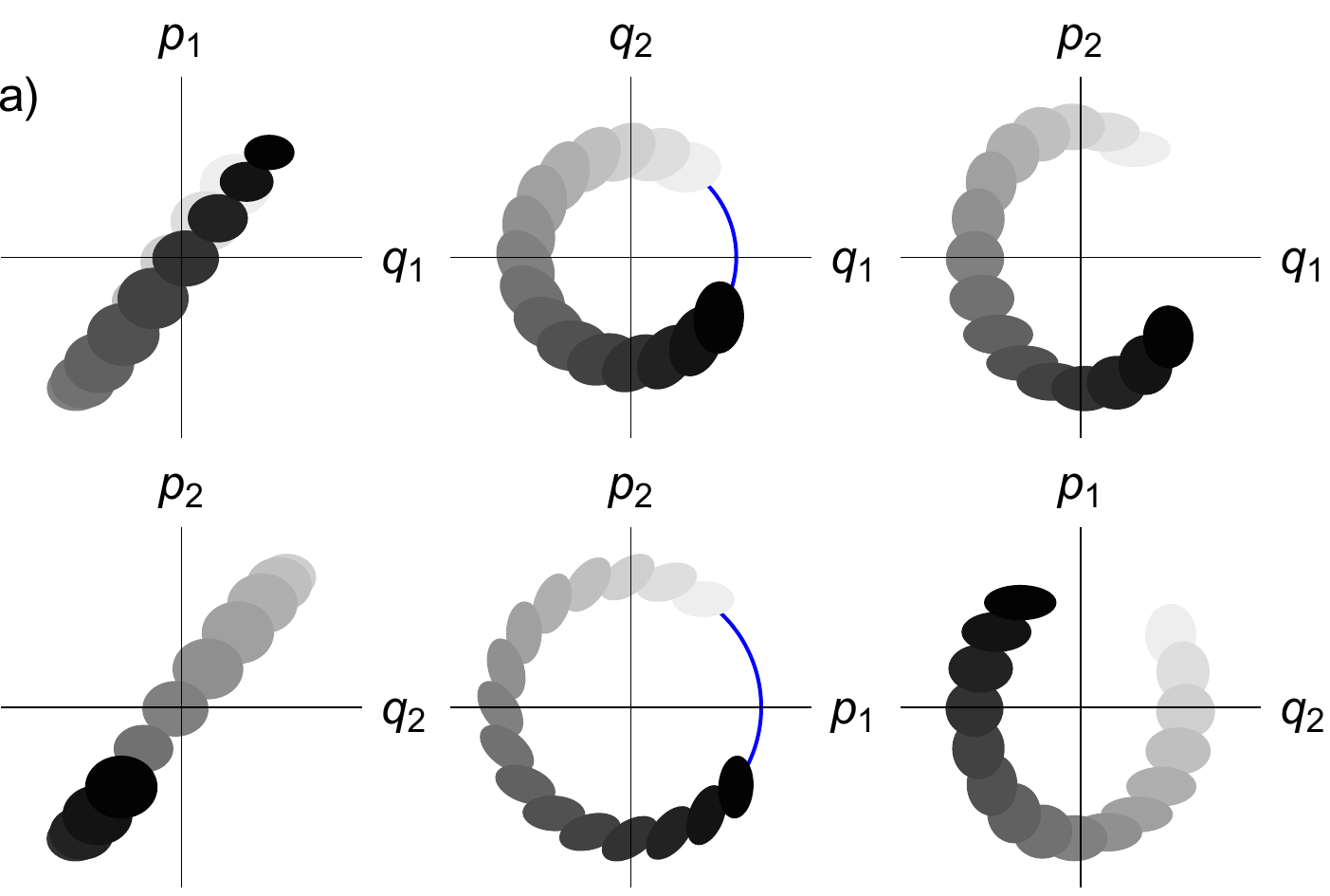}\label{Fig:JSPRot}
\vrule
\includegraphics[width=0.48\textwidth]{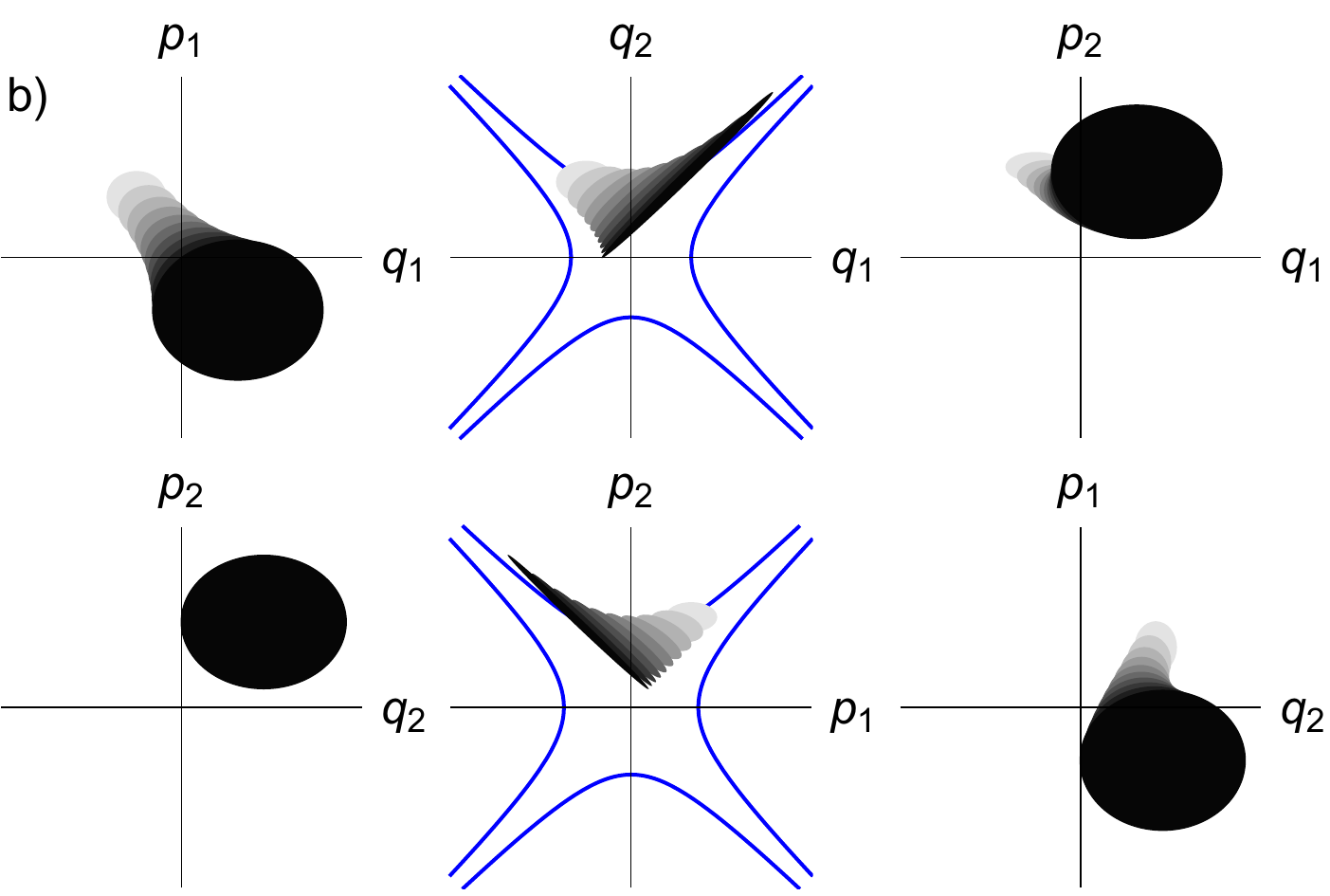}\label{Fig:JSASq}
\caption{
(Color online.) Two examples of multi-mode symplectic dynamics available to a pair of Gaussian modes. The dynamics in the four-dimensional phase space are here presented through six two-dimensional sections. The different shades of gray indicate time evolution: lighter ellipses correspond to earlier times. The dynamics depicted in these figures operate disjointedly in the $q_1 q_2$ and $p_1 p_2$. The path of the systems mean vector is plotted in blue. Equivalent dynamics are possible in the $q_1 p_2$ and $q_2 p_1$. \textbf{a)} Multi-mode rotation (section \ref{MultiModeRotation}). \textbf{b)} Multi-mode squeezing (section \ref{MultiModeSqueezing}).}\label{Fig:MultiModeDynamicsSymp}
\end{figure*}

Taking $N=2$ in equation \eqref{ASPDef} we find two types of multi-mode, symplectic, passive, and state-dependent dynamics. The first is given by
\bel{RotXXPP}
A_\textsc{sp}=\nu_{q_1q_2,p_1p_2}
\begin{pmatrix}
0 & \omega\\
\omega^\intercal & 0\\
\end{pmatrix}.
\ee
This dynamics acts disjointedly in the $q_1$, $q_2$ and $p_1$, $p_2$ planes.
The effect of this dynamics on the ellipsoid in phase space can be seen in Fig. \ref{Fig:MultiModeDynamicsSymp} a).
Within each of the $q_1$, $q_2$ and $p_1$, $p_2$ planes, this dynamics rotates the vector of means, $\bm{X}$, at a rate $\nu_{q_1q_2,p_1p_2}$. The ellipses in each of these planes also rotate at a rate $\nu_{q_1q_2,p_1p_2}$ about their centers.

It is critical to note that the rotations in each of these planes are either both clockwise or both  counter-clockwise. The Hamiltonian \eqref{HeffDef} associated with this dynamics is
\be
\hat{H}
=\nu_{q_1q_2,p_1p_2}(
\hat{q}_1\hat{p}_2
-\hat{p}_1\hat{q}_2)
+\text{H.c.}
\ee

The second type of multi-mode rotation is given by
\bel{RotXPXP}
A_\textsc{sp}=\nu_{q_1p_2,p_1q_2}
\begin{pmatrix}
0 & \openone_2\\
\openone_2 & 0\\
\end{pmatrix}.
\ee
This type of dynamics is identical to the previous dynamics, except now acting disjointedly in the $q_1$, $p_2$ and $p_1$, $q_2$ planes. The Hamiltonian \eqref{HeffDef} associated with this dynamics is
\be
\hat{H}
=\nu_{q_1p_2,p_1q_2}(
\hat{q}_1\hat{q}_2
+\hat{p}_1\hat{p}_2)
+\text{H.c.}
\ee

\subsubsection{Multi-mode Squeezing}\label{MultiModeSqueezing}
Taking $N=2$ in equation \eqref{ASADef}  we find two types of multi-mode, symplectic, active, and state-dependent dynamics. The first type of dynamics is given by,
\bel{SqXXPP}
A_\textsc{sa}
=\gamma_{q_1q_2,p_1p_2}
\begin{pmatrix}
0 & X\\
X & 0 \\
\end{pmatrix}.
\ee
This dynamics acts disjointedly in the $q_1$, $q_2$ and $p_1$, $p_2$ planes.
The effect of this dynamics on the ellipsoid in phase space can be seen in Fig. \ref{Fig:MultiModeDynamicsSymp} b). Within each of the $q_1$, $q_2$ and $p_1$, $p_2$ planes, this dynamics squeezes the state at a rate $\gamma_{q_1q_2,p_1p_2}$. The orientation of the squeezing is analogous to the $\times$-squeezing implemented by \eqref{x+Squeezing}.  In particular if the uncertainty in the $q_1+q_2$ direction is increased then the uncertainty in the $p_1+p_2$ direction is decreased. The Hamiltonian \eqref{HeffDef} associated with this dynamics is
\be
\hat{H}
=\gamma_{q_1q_2,p_1p_2}(\hat{q}_1\hat{p}_2-\hat{p}_1\hat{q}_2).
\ee

The second type of multi-mode squeezing is given by,
\bel{SqXPXP}
A_\textsc{sa}
=\gamma_{q_1p_2,p_1q_2}
\begin{pmatrix}
0 & Z\\
Z & 0 \\
\end{pmatrix}.
\ee
This type of dynamics is identical to the previous dynamics, except now acting disjointedly in the $q_1$, $p_2$ and $p_1$, $q_2$ planes. The Hamiltonian \eqref{HeffDef} associated with this dynamics is
\be
\hat{H}
=\gamma_{q_1p_2,p_1q_2}(\hat{q}_1\hat{q}_2+\hat{p}_1\hat{p}_2).
\ee

\begin{figure*}
\includegraphics[width=0.48\textwidth]{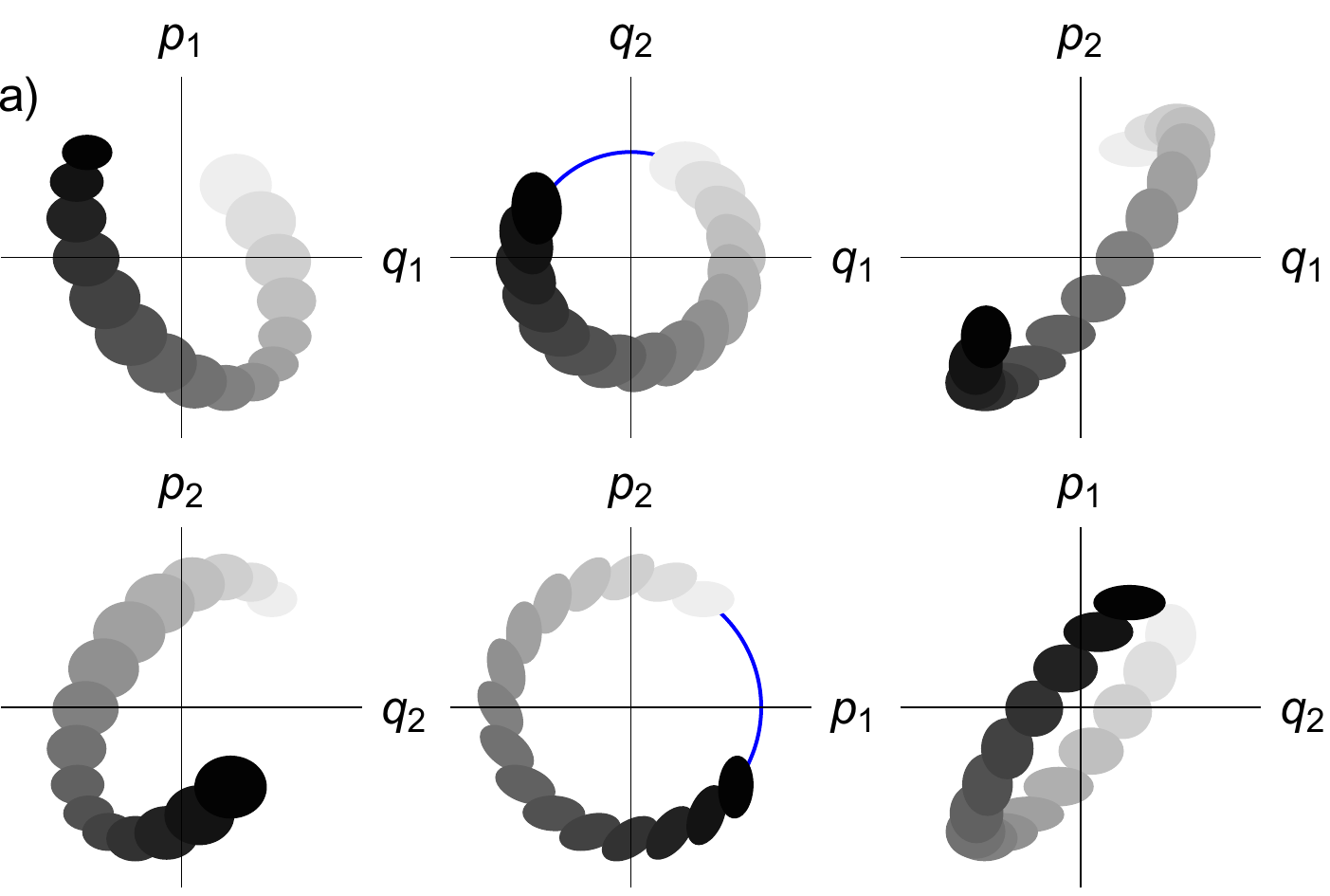}
\vrule
\includegraphics[width=0.48\textwidth]{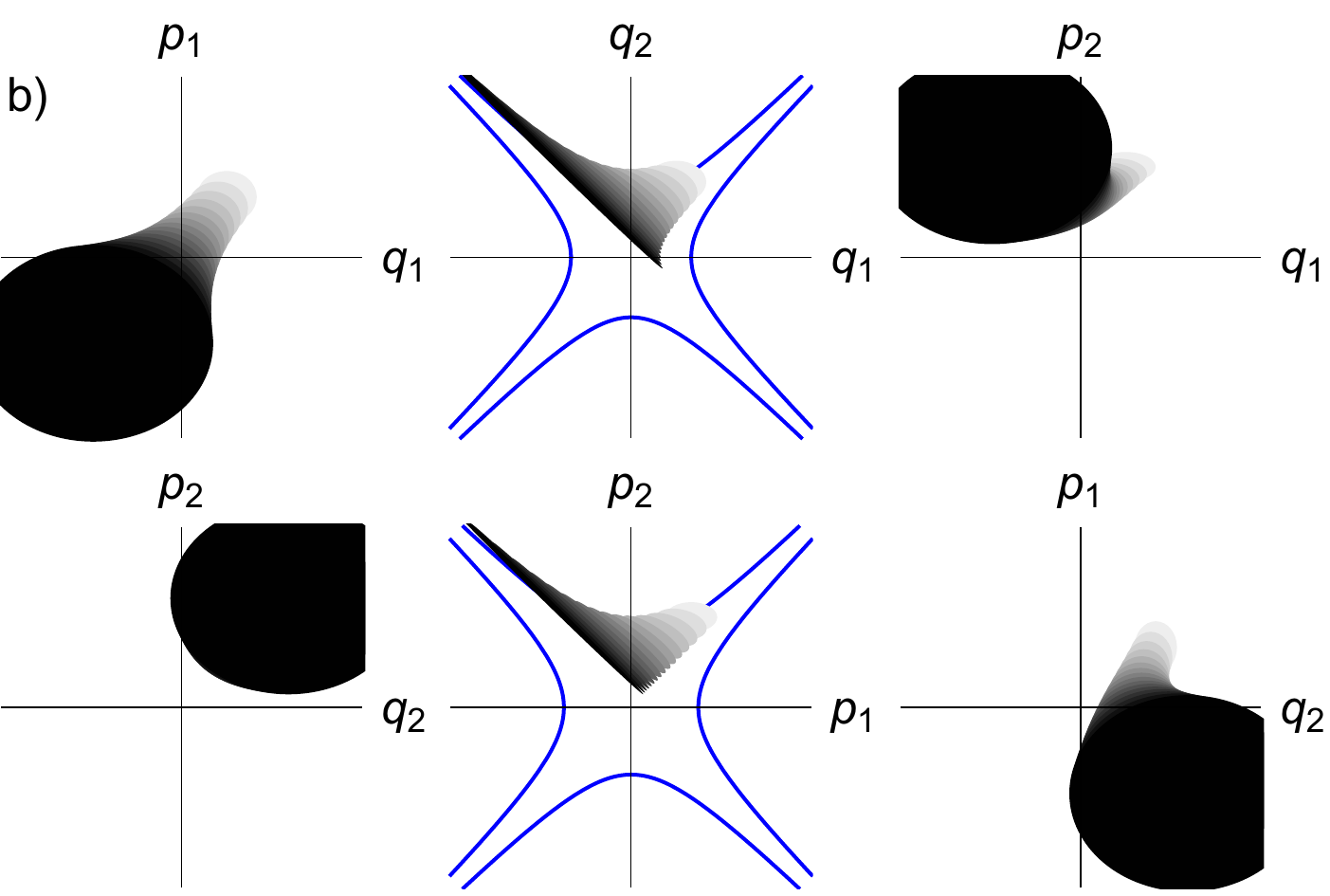}
\caption{
(Color online.) Two examples of multi-mode unsymplectic dynamics available to a pair of Gaussian modes. The dynamics in the four-dimensional phase space are here presented through six two-dimensional sections. The different shades of gray indicate time evolution: lighter ellipses correspond to earlier times. The dynamics depicted in these figures operate disjointedly in the $q_1 q_2$ and $p_1 p_2$ planes. The path of the systems mean vector is plotted in blue. Equivalent dynamics are possible in the $q_1 p_2$ and $q_2 p_1$. \textbf{a)} Multi-mode counter-rotation (section \ref{MultiModeCounterRotation}). \textbf{b)} Multi-mode counter-squeezing (section \ref{MultiModeCounterSqueezing}).
Neither of these dynamics are completely positive without the addition of a sufficiently large noise term. They are presented here without noise. The path of the systems mean vector is plotted in blue.}
\label{Fig:MultiModeDynamicsUns}
\end{figure*}

\subsubsection{Multi-mode Counter-Squeezing}\label{MultiModeCounterSqueezing}
Taking $N=2$ in equation \eqref{AUADef} we find two types of multi-mode, unsymplectic, active, and state-dependent dynamics. The first type of dynamics is given by
\bel{CSqXXPP}
A_\textsc{ua}
=\bar{\gamma}_{q_1q_2,p_1p_2}
\begin{pmatrix}
0 & \omega\\
\omega & 0
\end{pmatrix}.
\ee
This dynamics acts disjointedly in the $q_1$, $q_2$ and $p_1$, $p_2$ planes.  The effect of this dynamics on the ellipsoid in phase space can be seen in Fig. \ref{Fig:MultiModeDynamicsUns} b). Within each of the $q_1$, $q_2$ and $p_1$, $p_2$ planes, this dynamics squeezes the state at a rate $\bar{\gamma}_{q_1q_2,p_1p_2}$. The orientation of the squeezing is analogous to the $\times$-squeezing implemented by \eqref{x+Squeezing}. In particular if the uncertainty in the $q_1+q_2$ direction is increased then the uncertainty in the $p_1+p_2$ direction is also increased. The uncertainties in the orthogonal directions ($q_1-q_2$ and $p_1-p_2$) are therefore decreased. This is in contrast to the dynamics presented in Fig. \ref{Fig:MultiModeDynamicsSymp} b) in which the uncertainty in the $q_1+q_2$ and $p_1+p_2$ directions increase and decrease oppositely.

This dynamics is identical to the corresponding symplectic squeezing except that the squeezings happen in opposite directions, hence the name \textit{counter-squeezing}. A counter-squeezed state cannot be converted into a standardly squeezed state through a symplectic transformation. 

Even though this dynamics is unsymplectic (due to the counter-squeezing it does not preserve the symplectic form), it does preserve the volume of phase space. However, as we can see from Sec. \ref{GaussianCP}, a counter-squeezing alone cannot be completely positive, and would require a sufficient amount of free thermal noise (of the kind described in Sec. \ref{ThermalNoise}) to appear in completely positive dynamics. This thermal noise term will itself yield non-conservation of the volume of phase space, so it is correct to say that if we see counter-squeezing we also see non-conservation of phase space volume.

The second type of multi-mode counter-squeezing is given by
\bel{CSqXPXP}
A_\textsc{ua}
=\bar{\gamma}_{q_1p_2,p_1q_2}
\begin{pmatrix}
0 & \openone_2\\
-\openone_2 & 0
\end{pmatrix}.
\ee
This type of dynamics is identical to the previous dynamics, except now acting disjointedly in the $q_1$, $p_2$ and $p_1$, $q_2$ planes.

Just as before, this transformation is unsymplectic but preserves the volume of phase space. It is not completely positive on its own and requires a sufficient amount of free thermal noise. Thus, same as above, its appearance in completely positive dynamics implies that the overall dynamics will not preserve the volume of phase space.

\subsubsection{Multi-mode Counter-Rotation}\label{MultiModeCounterRotation}
Taking $N=2$ in equation \eqref{AUPDef} we find two types of multi-mode, unsymplectic, passive, and state-dependent dynamics. The first type of dynamics is given by
\bel{CRotXXPP}
A_\textsc{up}
=\bar{\nu}_{q_1q_2,p_1p_2}
\begin{pmatrix}
0 & X\\
-X & 0
\end{pmatrix}.
\ee
This type of dynamics acts disjointedly in the $q_1$, $q_2$ and $p_1$, $p_2$ planes. The effect of this dynamics on the ellipsoid in phase space can be seen in Fig. \ref{Fig:MultiModeDynamicsUns} a). Within each of the $q_1$, $q_2$ and $p_1$, $p_2$ planes, this dynamics rotates the vector of means, $\bm{X}$, at a rate $\bar{\nu}_{q_1q_2,p_1p_2}$. The ellipses in each of these planes also rotate at a rate $\bar{\nu}_{q_1q_2,p_1p_2}$ about their centers. It is critical to note that one of these rotations is clockwise while the other is counter-clockwise. This is in stark contrast with the dynamics described in Sec. \ref{MultiModeRotation} and displayed in Fig. \ref{Fig:MultiModeDynamicsSymp} a) in which the two planes rotate in the same direction.

This dynamics is identical to the corresponding symplectic rotation except that the rotation is in opposite directions hence the name \textit{counter-rotation}. As mentioned above (in Sec. \ref{MultiModeCounterSqueezing}), a counter-squeezed state cannot be converted into standardly squeezed state through a symplectic rotation. However, this transformation can be achieved using a counter-rotation. 

Note that, analogous to counter-squeezing, even though this dynamics is unsymplectic, it preserves the volume of phase space. Additionally note that this dynamics is passive, it admits no energy flow between the system and its environment. However, again, from Sec.\ref{GaussianCP} we can see that this process is not completely positive on its own and would require a sufficient amount of free thermal noise (of the kind described in Sec. \ref{ThermalNoise}). This noise will itself not preserve the volume of phase space and allow for energy flow between the system and its environment.

The second type of multi-mode counter-rotation is given by
\bel{CRotXPXP}
A_\textsc{up}
=\bar{\nu}_{q_1p_2,p_1q_2}
\begin{pmatrix}
0 & Z\\
-Z & 0
\end{pmatrix}.
\ee
This type of dynamics is identical to the previous dynamics, except now acting disjointedly in the $q_1$, $p_2$ and $p_1$, $q_2$ planes. 

Just as before, this transformation is unsymplectic but preserves the volume of phase space and is passive. However, it is not completely positive on its own and requires a sufficient amount of free thermal noise, which will itself be active and not preserve the volume of phase space.

\subsubsection{Multi-mode Squeezed}\label{MultiModeSqueezedNoise}
\begin{figure*}
\includegraphics[width=0.48\textwidth]{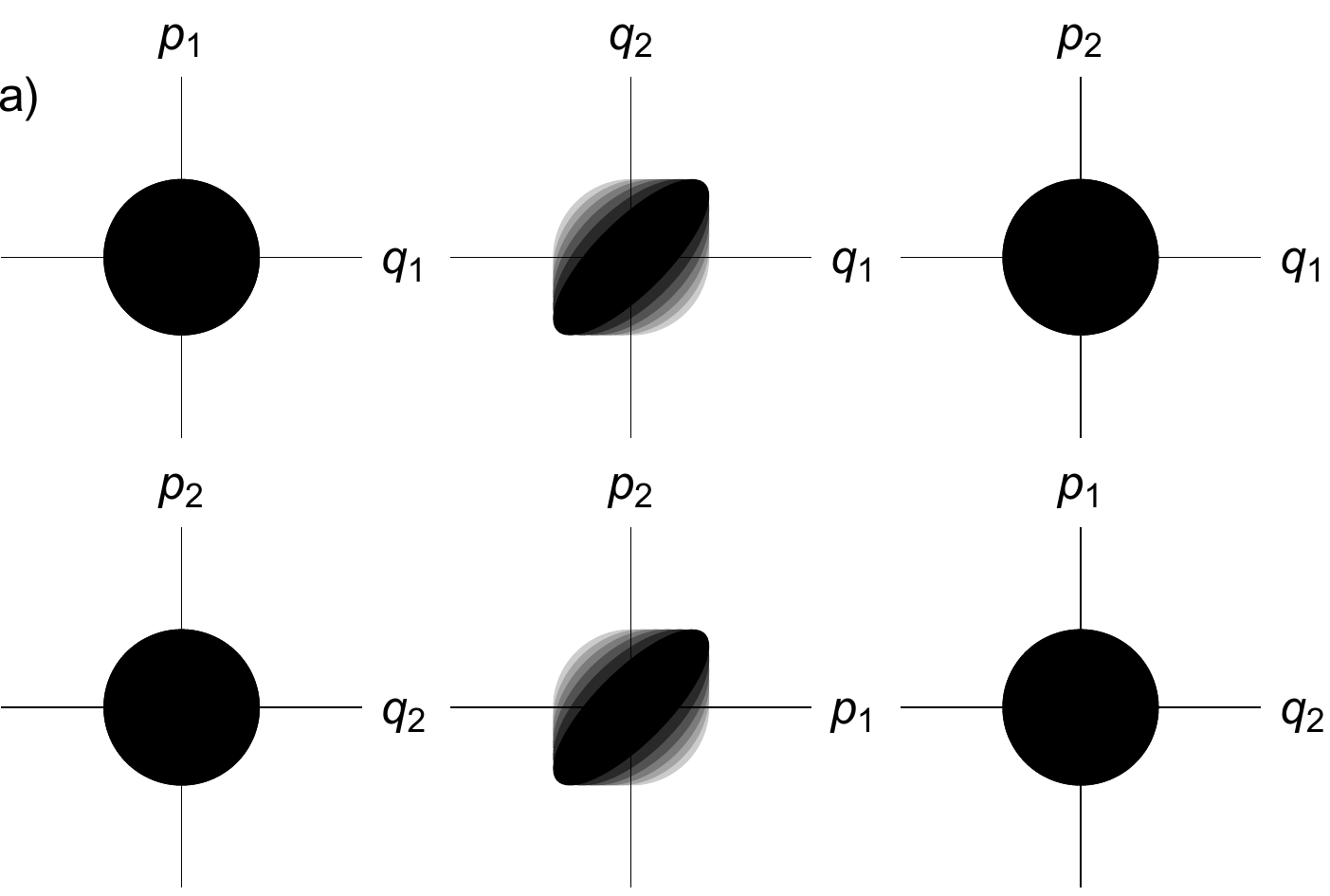}
\vrule
\includegraphics[width=0.48\textwidth]{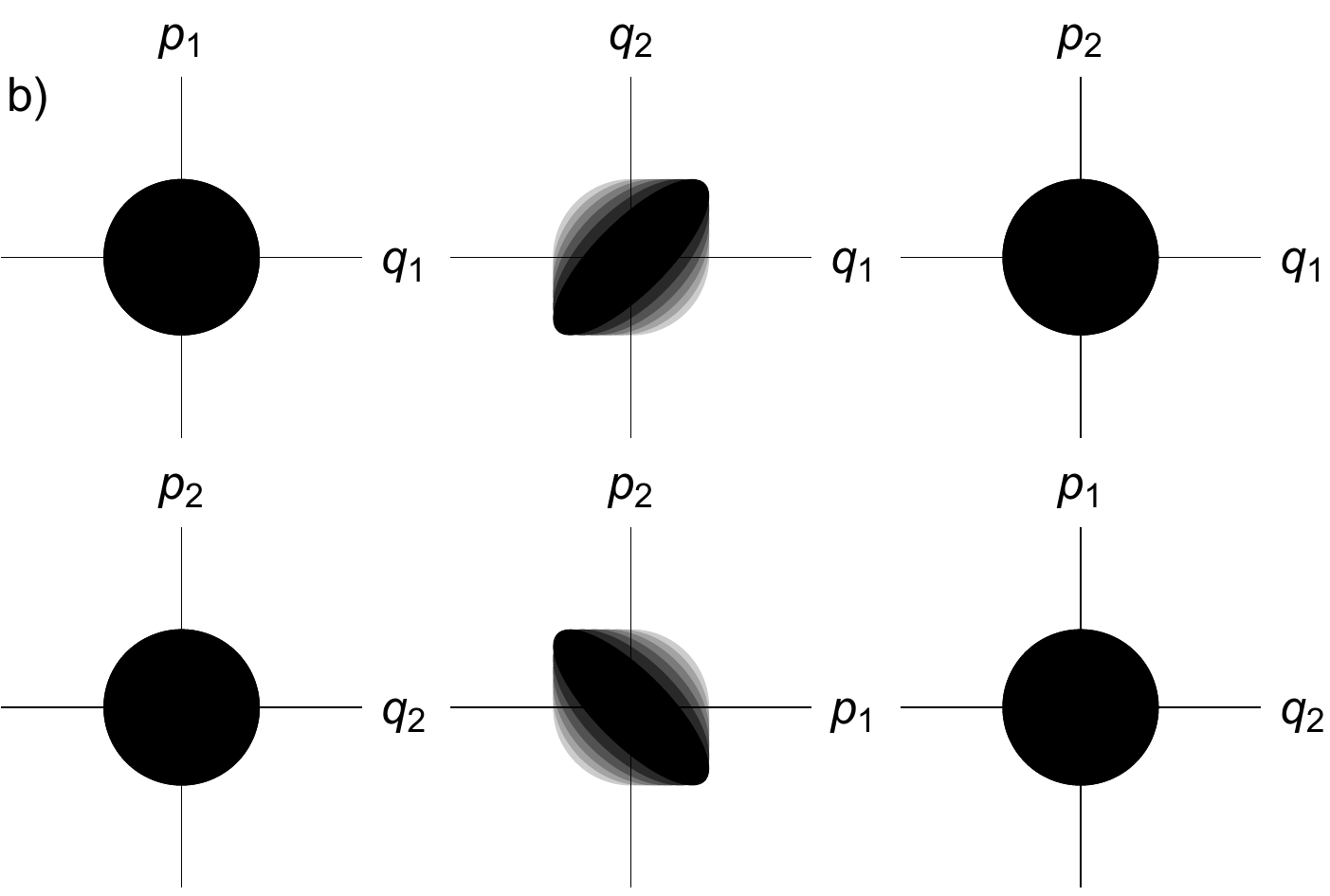}
\caption{
(Color online.) Two examples of anistropic noise in a pair of Gaussian modes. The dynamics in the four-dimensional phase space are here presented through six two-dimensional sections. The different shades of gray indicate time evolution: lighter ellipses correspond to earlier times. The dynamics depicted in these figures operate disjointedly in the $q_1 q_2$ and $p_1 p_2$ planes. Equivalent dynamics are possible in the $q_1 p_2$ and $q_2 p_1$. The cases displayed in \textbf{a)} and \textbf{b)} are those discussed individually in section \ref{MultiModeSqueezedNoise}.
}\label{Fig:MultiModeDynamicsNoise}
\end{figure*}
Taking $N=2$ in equation \eqref{CUPDef} we find four types of multi-mode, unsymplectic, passive, and state-independent dynamics. The first two types are given by
\be
C_\textsc{up}
=c_1\begin{pmatrix}
0 & \openone\\
\openone& 0
\end{pmatrix}
+c_z \begin{pmatrix}
0 & Z\\
Z & 0
\end{pmatrix}
\ee
for some real parameters $c_1$ and $c_z$. The effect the $c_1$ and $c_z$ dynamics have on the ellipse in phase space is shown in Fig. \ref{Fig:MultiModeDynamicsNoise} a) and b) respectively. This dynamics adds squeezed multi-mode noise to the system's covariance matrix while leaving the mean vector unchanged. Note that the covariance matrix only changes in the $q_1$, $q_2$ and $p_1$, $p_2$ planes.
Also note that in Fig. \ref{Fig:MultiModeDynamicsNoise} a) the $c_1$ dynamics increases the uncertainty in both the $q_1+q_2$ and $p_1+p_2$ directions whereas in Fig. \ref{Fig:MultiModeDynamicsNoise} b) the $c_z$ dynamics increases the uncertainty in one direction while decreasing the uncertainty in the other direction.

This dynamics is unsymplectic and does not preserve the volume of phase space. Recall that, as seen in Sec. \ref{GaussianCP}, such dynamics needs to be accompanied by a sufficient amount of free thermal noise (of the kind described in Sec. \ref{ThermalNoise}) in order to be completely positive.

The other two types of multi-mode squeezed noise are given by 
\be
C_\textsc{up}
=c_x \begin{pmatrix}
0 & X\\
X & 0
\end{pmatrix}
+c_w \begin{pmatrix}
0 & \omega\\
\omega^\intercal & 0
\end{pmatrix}.
\ee
These types of dynamics are identical to the previous dynamics, except now acting only in the $q_1$, $p_2$ and $p_1$, $q_2$ planes. Specifically the $c_x$ dynamics increases the uncertainty in both the $q_1+p_2$ and $p_1+q_2$ directions whereas the $c_w$ dynamics increases the uncertainty in one direction while decreasing the uncertainty in the other.

\section{Conclusions}
Taking a master equation approach to open Gaussian quantum mechanics, we developed a classification of the generators of open Gaussian dynamics. Specifically we divided the generators of Gaussian dynamics into: 1) active and passive, 2) symplectic and non-symplectic, 3) single-mode and multi-mode, and 4) state-dependent and state-independent. Visualizations with detailed descriptions were provided for each type of dynamics.

Through this classification we analyzed the relationships between the differents parts of the dynamics  and the role of noise in the context of complete positivity. Specifically we found that non-symplectic dynamics requires noise in order to be completely positive, and this noise makes any completely positive non-symplectic dynamics necessarily non-passive. Additionally since noise does not conserve the volume of phase space, completely positive non-symplectic dynamics does not conserve the volume of phase space. 

We note that it is not the case that non-symplectic transformations automatically violate the conservation of phase space volume. Whether or not they do depends critically on whether or not they are completely positive. In fact, there can be non-CP processes (e.g. non-Markovian dynamics) which do not conserve the symplectic form while conserving the volume of phase space.

We also discussed the consequences of this on the relationship between information and energy flows in open quantum mechanics. We found that Gaussian dynamics which generates entanglement cannot be guaranteed to have zero energy flow for all states. Generically if entanglement is generated with the environment there will be system-environment energy flow as well.

Work that applies the new results to the dynamics of quantum systems that are bombarded by a rapid succession of ancillae is in progress.

\acknowledgments

AK, EMM and RBM acknowledge support through the Discovery program of the Natural Sciences and Engineering Research Council of Canada (NSERC). EMM acknwoledge support of the ONtario Early Researcher Award. EGB also acknowledges support by NSERC through their Postdoctoral Fellowship. DG acknowledges support from Vanier Canada Graduate Scholarships (CGS) and help from Ben Grimmer on the complete positivity proof in Appendix \ref{AppCPProof}.

\appendix
\section{Gaussian Quantum Mechanics}\label{AppGQM}
\subsection{Standardizing a Quadratic Hamiltonian}
In this appendix we cast a generic quadratic Hamiltonian in a standard form. A quadratic Hamiltonian is an operator which can be written in the form
\begin{align}
\hat{H}
&=\frac{1}{2}\hat{\bm{X}}^\intercal  \, F \, \hat{\bm{X}}
+\bm{\alpha}^\intercal\hat{\bm{X}}
+\beta \, \hat{\openone}\\
&\nonumber
=\frac{1}{2}F_j{}^k \, \hat{X}^j \, \hat{X}_k
+\bm{\alpha}^i\hat{X}_i
+\beta \, \hat{\openone}
\end{align}
with the sole restriction that it is . 

In order to analyze what this restriction means for the coefficients $F$, $\alpha$, and $\beta$, we need to cast the Hamiltonian as a sum of linearly independent terms. Each of the $\hat{X}_i$ are linearly independent from each other and from all the other terms. Note however that quadratic and constant terms, $\hat{X}^j \, \hat{X}_k$ and $\hat{\openone}$, are not linearly independent since we have the commutation relation
\be
\ii \Omega_k{}^j \, \hat{\openone}
=\hat{X}^j \, \hat{X}_k -\hat{X}_k \, \hat{X}^j.
\ee
More generally, any antisymmetric sum of the quadrature operators will be proportional to the identity. Ultimately this introduces an ambiguity when one tries to identify a Hamiltonian in terms of its coefficients.

We can resolve this ambiguity by converting rewriting the antisymmetric parts of $F$ as a part of $\beta$. Breaking $F$ into its symmetric and antisymmetric parts as $F=F_\text{sym}+ F_\text{anti}$, we can compute the term in the Hamiltonian coming from $F_\text{anti}$ as 
\begin{align}
\frac{1}{2}\hat{\bm{X}}^\intercal F_\text{anti} \, \hat{\bm{X}}
&=\frac{1}{2} (F_\text{anti})_j{}^k \, \hat{X}^j \, \hat{X}_k\\
&\nonumber
= \frac{1}{4}((F_\text{anti})_j{}^k-(F_\text{anti})^k{}_j) \, \hat{X}^j \, \hat{X}_k\\
&\nonumber
= \frac{1}{4}(F_\text{anti})_j{}^k \, (\hat{X}^j \, \hat{X}_k -\hat{X}_k \, \hat{X}^j)\\
&\nonumber
= \frac{\ii}{4}(F_\text{anti})_j{}^k \, 
\Omega^j{}_k \, \hat{\openone}.
\end{align}
We can absorb this into the $\beta$ term by taking
\be
\beta\to\beta+\frac{\ii}{4}(F_\text{anti})_j{}^k \, \Omega^j{}_k \, .
\ee
Thus without loss of generality we can take $F$ to be symmetric. Thus we can write any quadratic Hamiltonian in the standardized form
\begin{align}
\hat{H}
&=\frac{1}{2}\hat{\bm{X}}^\intercal  \, F \, \hat{\bm{X}}
+\bm{\alpha}^\intercal\hat{\bm{X}}
+\beta \, \hat{\openone}\\
&\nonumber
=\frac{1}{2}F_j{}^k \, \hat{X}^j \, \hat{X}_k
+\bm{\alpha}^i\hat{X}_i
+\beta \, \hat{\openone}
\end{align}
with $F$ symmetric. 

Now that the Hamiltonian is written as the sum of linearly independent terms, we can impose the restriction that $\hat{H}$ is Hermitian, $\hat{H}^\dagger=\hat{H}$. Specifically we do this by computing $\hat{H}^\dagger$ and requiring its coefficients to be identical to those of $\hat{H}$. We find
\begin{align}
\hat{H}^\dagger
&=\big(\frac{1}{2}F_j{}^k \, \hat{X}^j \, \hat{X}_k
+\bm{\alpha}^i\hat{X}_i
+\beta \, \hat{\openone}\big)^\dagger\\
&\nonumber
=\frac{1}{2}F_j{}^k{}^* \, \hat{X}_k \, \hat{X}^j
+\bm{\alpha}^i{}^*\hat{X}_i
+\beta^* \, \hat{\openone}\\
&\nonumber
=\frac{1}{2}(F^\dagger)^k{}_j \, \hat{X}_k \, \hat{X}^j
+(\bm{\alpha}^*)^i\hat{X}_i
+\beta^* \, \hat{\openone}\\
&\nonumber
=\frac{1}{2}(F^\dagger)_k{}^j \, \hat{X}^k \, \hat{X}_j
+(\bm{\alpha}^*)^i\hat{X}_i
+\beta^* \, \hat{\openone}
\end{align}
This allows us to conclude that $F$ is Hermitian ($F^\dagger=F$), that $\bm{\alpha}$ is real ($\bm{\alpha}^*=\bm{\alpha}$), and that $\beta$ is real ($\beta^*=\beta$). Since $F$ was already symmetric, it being Hermitian implies that it is also real.

Thus we can always take any quadratic Hamiltonian to be of the form
\begin{align}
\hat{H}
&=\frac{1}{2}\hat{\bm{X}}^\intercal  \, F \, \hat{\bm{X}}
+\bm{\alpha}^\intercal\hat{\bm{X}}
+\beta \, \hat{\openone}\\
&\nonumber
=\frac{1}{2}F_j{}^k \, \hat{X}^j \, \hat{X}_k
+\bm{\alpha}^i\hat{X}_i
+\beta \, \hat{\openone}
\end{align}
with $F$ real and symmetric, $\bm{\alpha}$ real, and $\beta$ real. Furthermore we can drop the $\beta$ term as it only provides a constant energy offset to the dynamics. Thus we have the form claimed in \eqref{QuadHamForm}.

\subsection{Derivation of Gaussian Unitary Master Equation}
In this appendix we translate the unitary master equation with a quadratic Hamiltonian from Hilbert space to phase space. 

As shown above any quadratic Hamiltonian can be written in the form,
\be
\hat{H}=\frac{1}{2}\hat{\bm{X}}^\intercal  \, F \, \hat{\bm{X}}
+\bm{\alpha}^\intercal\hat{\bm{X}}
\ee
where $F$ is real and symmetric, and $\bm{\alpha}$ is real.

Taking the Heisenberg picture (with $\hbar=1$), the components of the operator vector, $\bm{X}$, evolve as
\begin{align}
&\frac{\d}{\d t}\hat{X}_r
=\ii \, [\hat{H},\hat{X}_r]\\
\nonumber
&=\frac{\ii}{2} \, F_j{}^k [\hat{X}^j \hat{X}_k,\hat{X}_r]
+\ii \, \alpha^{i} [\hat{X}_i,\hat{X}_r]\\
\nonumber
&=\frac{\ii}{2} F_j{}^k\! \hat{X}^j[\hat{X}_k,\!\hat{X}_r]
+\!\frac{\ii}{2} F_j{}^k[\hat{X}^j,\!\hat{X}_r]\hat{X}_k
+\!\ii \, \alpha^{i} [\hat{X}_i,\!\hat{X}_r]\\
\nonumber
&=\frac{-1}{2}F_j{}^k \, \Omega_{kr} \, \hat{X}^j
-\frac{1}{2}F_j{}^k \, \Omega^j{}_r \, \hat{X}_k
-\alpha^{i} \, \Omega_{ir} \, \hat{\openone}\\
\nonumber
&=\frac{-1}{2}F^j{}_k \, \Omega^k{}_r \, \hat{X}_j
-\frac{1}{2}F_k{}^j \, \Omega^k{}_r \, \hat{X}_j
-\alpha_k \, \Omega^k{}_r \, \hat{\openone}\\
\nonumber
&=-\Omega^k{}_r\big(\frac{1}{2}(F_k{}^j+F^j{}_k) \, \hat{X}_j
+\alpha_k\, \hat{\openone}\big)\\
\nonumber
&=\Omega_r{}^k\big(F_k{}^j \, \hat{X}_j
+\alpha_k \, \hat{\openone}\big).
\end{align}
Thus we have the operator vector evolving as,
\be
\frac{\d}{\d t}\hat{\bm{X}}
=\Omega (F \hat{\bm{X}}+\alpha \hat{\openone}).
\ee
Thus unitary evolution in the Hilbert space corresponds to linear-affine evolution in the phase space.

\subsection{Solving the time independent gaussian unitary master equation}
In this appendix we solve the time independent unitary master equation derived in the previous appendix,
\bel{AppUnitaryMasterEq}
\frac{\d}{\d t}\hat{\bm{X}}
=\Omega (F \hat{\bm{X}}+\alpha \hat{\openone}).
\ee
In order to do this we embed $\hat{\bm{X}}$ into an affine space. This will have the effect of linearizing the master equation. 

We define the $2N+1$ dimensional operator-valued vector $\hat{\bm{Y}}=(\hat{\openone}, \hat{\bm{X}}^\intercal)^\intercal$. From \eqref{AppUnitaryMasterEq} we can see that this new vector evolves as
\bel{yProp}
\frac{\d\hat{\bm{Y}}}{\d t}
=\begin{pmatrix}
0 & 0\\
\Omega\bm{\alpha} & \Omega F
\end{pmatrix}
\hat{\bm{Y}}.
\ee
Note that the top row of $\hat{\bm{Y}}$'s propagator is zero because unitary evolution is unital, that is it preserves the identity operator.

Since \eqref{yProp} is now a linear differential equation we can quickly solve it as,
\begin{align}
\hat{\bm{Y}}(t)
&=\text{exp}\Bigg(
\begin{pmatrix}
0 & 0\\
\Omega\bm{\alpha} & \Omega F
\end{pmatrix}
t\Bigg) \, \hat{\bm{Y}}(0)\\
&\nonumber
=\begin{pmatrix}
1 & 0\\
\frac{\text{exp}(\Omega F \, t)-\openone_{2N}}{\Omega F} \, \Omega\bm{\alpha} & \text{exp}(\Omega F \, t)
\end{pmatrix}
\hat{\bm{Y}}(0).
\end{align}
Taking
\begin{align}
S(t)&=\text{exp}(\Omega F \, t)\\ 
\bm{d}(t)&=\frac{\text{exp}(\Omega F \, t)-\openone_{2N}}{\Omega F} \, \Omega\bm{\alpha}
\end{align}
yields
\begin{align}
\hat{\bm{Y}}(t)
&=\begin{pmatrix}
\hat{\openone}\\
\hat{\bm{X}}(t)
\end{pmatrix}
=\begin{pmatrix}
1 & 0\\
\bm{d}(t) \, & S(t)
\end{pmatrix}
\begin{pmatrix}
\hat{\openone}\\
\hat{\bm{X}}(0)
\end{pmatrix}
\end{align}
such that,
\be
\hat{\bm{X}}(t)=S(t)\hat{\bm{X}}(0)+\bm{d}(t)\hat{\openone}.
\ee
Thus we have identified $S(t)$ and $\bm{d}(t)$ to match the expressions given by \eqref{SHamDef} and \eqref{dHamDef}.

\subsection{Form of a General Open Update}
In this appendix we derive the form of a general open Gaussian channel given the form of the unitary channel, \eqref{SymplecticXUp} and \eqref{SymplecticVUp}.

Analogously to the Stinespring dilation theorem, we can, without loss of generality, take our channel to be a unitary channel in a larger phase space with an ancilla that is initially uncorrelated with our system. We thus need to review the process for adding and removing degrees of freedom to phase space.

Concretely, given observables $\hat{\bm{X}}_S$ and $\hat{\bm{X}}_A$ for our system and ancilla respectively, their joint operators are
\bel{JointOperatorVectorDef}
\hat{\bm{X}}_{SA}
\coloneqq\hat{\bm{X}}_S\oplus\hat{\bm{X}}_A
=(\hat{\bm{X}}_S,\hat{\bm{X}}_A)^\intercal.
\ee
Since the systems are independent, their joint symplectic form is
\be
\Omega_{SA}\coloneqq
\Omega_S\oplus\Omega_A
=\begin{pmatrix}
\Omega_S & 0\\
0 & \Omega_A
\end{pmatrix}.
\ee
From \eqref{JointOperatorVectorDef} we can define the joint mean vector and joint covariance matrix, $\bm{X}_{SA}$ and $\sigma_{SA}$. Explicitly one finds
\be
\bm{X}_{SA}=(\bm{X}_S,\bm{X}_A)^\intercal
\ee
and
\be
\sigma_{SA}=
\begin{pmatrix}
\sigma_S & \gamma\\
\gamma^\intercal & \sigma_A
\end{pmatrix}
\ee
where $\bm{X}_S$, $\bm{X}_A$, $\sigma_S$, and $\sigma_A$ capture the reduced states of the system and ancilla while $\gamma$ captures the correlations between them.

As discussed above any Gaussian update of $X_\text{S}$ and $\sigma_\text{S}$ can be viewed as a symplectic-affine transformation in a larger phase space with an initially uncorrelated ancilla, here meaning $\gamma(0)=0$. Working in such a larger phase space we can find the generic form for an update of $X_\text{S}$ and $\sigma_\text{S}$.  By assumption, the joint system undergoes a symplectic-affine transformation,
\bel{JointSympXUp}
\bm{X}_{SA}
=S_{SA} \ \bm{X}_{SA}(0)
+d_{SA}
\ee
and 
\bel{JointSympVUp}
\sigma_{SA}
=S_{SA} \, \sigma(0) \, S_{SA}^\intercal.
\ee
We can explicitly decompose $S_{SA}$ and $d_{SA}$ over the tensor sum by dividing them into subblocks as
\begin{align}
S_{SA}
&=\begin{pmatrix}
T_S & M\\
L & T_A
\end{pmatrix}
\end{align}
for some $T_S$, $M$, $L$, and $T_A$ and 
\be
\bm{d}_{SA}
=\begin{pmatrix}
\bm{d}_{S}\\
\bm{d}_{A}
\end{pmatrix}
\ee
for some $\bm{d}_{S}$ and $\bm{d}_{A}$.

Working out \eqref{JointSympXUp} and \eqref{JointSympVUp} in terms of these subblocks yield
\begin{align}
\bm{X}_S
&=T_S \, \bm{X}_S(0)
+M \, \bm{X}_A(0)
+\bm{d}_{S}\\
\bm{X}_A
&=T_A \, \bm{X}_A(0)
+L \, \bm{X}_S(0)
+\bm{d}_{A}\\
\sigma_S
&=T_S \, \sigma_S(0) \, T_S^\intercal
+M \, \sigma_A(0) \, M^\intercal\\
\gamma
&=T_S \, \sigma_S(0) \, M^\intercal
+L \, \sigma_A(0) \, T_A^\intercal\\
\sigma_A
&=L \, \sigma_S(0) \, L^\intercal
+T_A \, \sigma_A(0) \, T_A^\intercal.
\end{align}
Note that we have used $\gamma(0)=0$.

Defining
\begin{align}
T&\coloneqq T_S\\
\bm{d}&\coloneqq M \ \bm{X}_A(0)
+\bm{d}_{S}\\
R&\coloneqq M \, \sigma_A(0) \, M^\intercal
\end{align}
we find the general update on $X_\text{S}$ and $\sigma_S$ to be of the form
\begin{align}
\bm{X}_S
&=T \, \bm{X}_S(0)+\bm{d}\\
\sigma_S
&=T \, \sigma_S(0) \, T^\intercal+R.
\end{align}
Matching the form claimed in equations \eqref{GeneralUpdateX} and \eqref{GeneralUpdateV}.

Note that since $S_{SA}$ and $\bm{d}_{SA}$ were real valued, so are $T$, $\bm{d}$, and $R$. Moreover $R$ is manifestly symmetric.

\subsection{Derivation of Complete Positivity Condition}\label{AppCPProof}
In this appendix we show that a Gaussian update of the form given by \eqref{GeneralUpdateX} and \eqref{GeneralUpdateV}, i.e.,
\begin{align}
\bm{X}
&=T \, \bm{X}(0)+\bm{d}\\
\label{AppSigmaUpdate}
\sigma
&=T \, \sigma(0) \, T^\intercal+R.
\end{align}
is completely positive if and only if 
\bel{AppFiniteCPCond}
R+\ii \, \Omega-\ii \, T \, \Omega \,  T^\intercal\geq0.
\ee
Note that \eqref{AppFiniteCPCond} is equivalent to the condition, 
\bel{AppFiniteCPCondAlt}
R-\ii \, \Omega+\ii \, T \, \Omega \,  T^\intercal\geq0
\ee
since $R$, $T$, and $\Omega$ are real valued and complex conjugation preserves positive semidefiniteness.

First we will establish that satisfying \eqref{AppFiniteCPCond} is necessary and sufficient for Gaussian dynamics to be positive, i.e. that it maps valid states to valid states. Specifically this means,
\bel{AppPresPos}
\sigma(0)\geq\ii \, \Omega \implies \sigma=T \, \sigma(0) \, T^\intercal+R\geq\ii \, \Omega
\ee
for every real symmetric $\sigma(0)$. Then we extend this argument to entangled Gaussian states showing the same condition is necessary for complete positivity.

To show \eqref{AppFiniteCPCond} is necessary and sufficient for positive dynamics we begin by rearranging the expression for the validity of the updated state
\bel{AppAlgebra0}
\sigma=T \, \sigma(0) \, T^\intercal+R
\geq\ii \, \Omega.
\ee
Adding $\ii \, T \, \Omega \, T^\intercal$ to each side of \eqref{AppAlgebra0} and rearranging terms yields,
\bel{AppAlgebra1}
R-\ii \, \Omega+\ii \, T \, \Omega \,  T^\intercal
\geq-T(\sigma(0)-\ii\,\Omega)T^\intercal.
\ee
Note that since the initial state $\sigma(0)$ is a valid state, i.e. $\sigma(0)\geq\ii\,\Omega$, the right hand side of \eqref{AppAlgebra1} is always negative semidefinite. 

Thus if we assume that our update obeys \eqref{AppFiniteCPCondAlt} then we automatically have
\be
R-\ii \, \Omega+\ii \, T \, \Omega \,  T^\intercal
\geq0
\geq-T(\sigma(0)-\ii\,\Omega)T^\intercal.
\ee
Therefore \eqref{AppFiniteCPCondAlt} is sufficient to guarantee the positivity of the update.

To show why \eqref{AppFiniteCPCondAlt} is also necessary, we suppose that  \eqref{AppFiniteCPCondAlt} does not hold, such that for some complex unit vector $\bm{v}$, we have 
\be
-\lambda=\bm{v}^\dagger\big(R-\ii \, \Omega+\ii \, T \, \Omega \,  T^\intercal\big)\bm{v}<0.
\ee
If we can find a real symmetric matrix $\sigma(0)$ with \mbox{$\sigma(0)\geq\ii\,\Omega$} such that
\be
\bm{v}^{*}T \, (\sigma(0)-\ii\,\Omega) \, T^\intercal\bm{v}<\lambda
\ee
then we will have constructed a violation of \eqref{AppPresPos}.

Suppose instead that we are tasked to find a positive semidefinite matrix $M\geq0$ such that
\be
\bm{v}^{*}T \, M \, T^\intercal\bm{v}<\lambda.
\ee
This is a trivial task as we can just take $M$ to have an eigenvector $\bm{u}=T^\intercal\bm{v}$ with an arbitrarily small eigenvalue. From this we can nearly construct the state $\sigma(0)$ desired above. We do this by breaking $M$ into its real and imaginary parts as
\begin{align}
M&=\sigma_M-\ii\Omega_M\\
\sigma_M&\coloneqq\frac{1}{2}(M^\dagger+M)\\
\Omega_M&\coloneqq\frac{1}{2\ii}(M^\dagger-M)
\end{align}
and noting that since $M$ is Hermitian we automatically have that $\sigma_M$ is symmetric and $\Omega_M$ is antisymmetric. Additionally since $M\geq0$ we have $\sigma_M\geq\ii\Omega_M$. If we can find an $M$ such that $\Omega_M=\Omega$ then  taking $\sigma(0)=\sigma_M$ completes the necessity proof. Unfortunately we were unable to construct such an $M$, but we none the less claim it can be done. Withstanding this gap, this completes the proof of necessity.

Above we have established the necessity and sufficiency of \eqref{AppFiniteCPCondAlt} for positivity of the dynamics, i.e. that the transformation takes valid states to valid states. In general however it is not enough to just require positivity, one must also impose complete positivity, i.e. that extending the map to act trivially on an additional spectator system does not spoil its positivity.

In Gaussian Quantum mechanics, additional systems are added via a direct sum. Taking our system S to have N modes and the ancilla to have M modes. The symplectic form for a bipartite system is
\be
\Omega_{SA}
=\Omega_S\oplus\Omega_A
=\begin{pmatrix}
\Omega_S & 0\\
0 & \Omega_A
\end{pmatrix}
\ee
A transformation which acts trivially on the ancillary system is of the form,
\begin{align}
T_{SA}
&=T_S\oplus\openone_{2 M}
=\begin{pmatrix}
T_S & 0\\
0 & \openone_{2 M}
\end{pmatrix}\\
\bm{d}_{SA}
&=\bm{d}_S\oplus\bm{0}
=\begin{pmatrix}
\bm{d}_S\\
\bm{0}
\end{pmatrix}\\
R_{SA}
&=R_S\oplus 0
=\begin{pmatrix}
R_S & 0\\
0 & 0
\end{pmatrix}.
\end{align}
The transformation given by $T_S$, $\bm{d}_S$ and $R_S$ is completely positive if and only if the transformation given by $T_{SA}$, $\bm{d}_{SA}$ and $R_{SA}$ is positive, i.e. 
\be
R_{SA}+\ii \, \Omega_{SA}-\ii \, T_{SA} \, \Omega_{SA} \,  T_{SA}^\intercal\geq0.
\ee
Straightforward computation yields
\begin{align}
R_{SA}+\ii \, \Omega_{SA}-\ii \, T_{SA} \, \Omega_{SA} \,  T_{SA}^\intercal\\
=\begin{pmatrix}
R_{S}+\ii \, \Omega_{S}-\ii \, T_{S} \, \Omega_{S} \,  T_{S}^\intercal & 0\\
0 & 0
\end{pmatrix}    
\end{align}
such that for Gaussian dynamics, positivity is equivalent to complete positivity.

\subsection{Solving the time independent open gaussian master equation}
In this appendix we solve the general Gaussian master equation \eqref{GeneralDiffXUp} and \eqref{GeneralDiffVUp},
\begin{align}
\label{AppGeneralDiffXUp}
\frac{\d}{\d t}\bm{X}(t)
&=\Omega(A\bm{X}(t)+\bm{b})\\
\label{AppGeneralDiffVUp}
\frac{\d}{\d t}\sigma(t)
&=(\Omega A) \, \sigma(t)
+\sigma(t) \, (\Omega A)^\intercal
+C.
\end{align}
This amounts to finding $T$, $\bm{d}$, $R$ in equations \eqref{GeneralUpdateX} and \eqref{GeneralUpdateV} in terms of $A$, $\bm{b}$, and $C$.

The first of these equations is formally identical to \eqref{AppUnitaryMasterEq} and can be solved using the method laid out in that appendix. That is, by embedding $\bm{X}$ into an affine space we can convert the linear-affine equation \eqref{AppGeneralDiffXUp} into a linear one which is quickly solved. This yields,
\begin{align}
T(t)&=\exp(\Omega A  \, t)\\
\bm{d}(t)&=\frac{\exp(\Omega A \,  t)-\openone_{2N}}{\Omega A}\bm{b}.
\end{align}
as claimed in \eqref{TdRExplicit}.

This still leaves us the challenge of writing $R(t)$ in terms of $A$ and $C$. Note that \eqref{AppGeneralDiffVUp} is still linear-affine, only now it is a matrix equation. We can still use the embedding in an affine space technique but we must first covert the matrix $\sigma$ into a vector. We do this using the $\vec$ operation, which forms a vector from a matrix by listing its entries. For instance,
\be
\vec\begin{pmatrix}
a & b \\
c & d
\end{pmatrix}
=(a,b,c,d)^\intercal.
\ee
It is helpful the way that $\vec$ acts on an outer product. One can quickly confirm that for any vectors $\bm{x}$ and $\bm{y}$,
\be
\vec(\bm{x}\bm{y}^\intercal)
=\bm{x}\otimes\bm{y}.
\ee
From this we can see that, under $\vec$, operating on the left of an outer product becomes operating on the first tensor factors. For any linear operator $A$
\begin{align}
\vec(A\bm{x}\bm{y}^\intercal)
&=\vec\big((A\bm{x})\bm{y}^\intercal\big)\\
&\nonumber
=(A\bm{x})\otimes\bm{y}\\
&\nonumber
=(A\otimes\boldsymbol{1})\bm{x}\otimes\bm{y}\\
&\nonumber
=(A\otimes\boldsymbol{1}) \, \vec(\bm{x}\bm{y}^\intercal).
\end{align}
Likewise operating on the right of an outer product becomes adjoint operating on the second factors space. For any linear operator $B$
\begin{align}
\vec(\bm{x}\bm{y}^\intercal B)
&=\vec\big(\bm{x}(B^\intercal \bm{y})^\intercal\big)\\
&\nonumber
=\bm{x}\otimes(B^\intercal \bm{y})\\
&\nonumber
=(\boldsymbol{1}\otimes B^\intercal) \bm{x}\otimes\bm{y}\\
&\nonumber
=(\boldsymbol{1}\otimes B^\intercal) \, \vec(\bm{x}\bm{y}^\intercal).
\end{align}
Since outer products span the matrices, by linearity, we have as claimed in \eqref{vecIdentity},
\be
\vec(ACB^\intercal)=(A\otimes B)\,\vec(C)
\ee
for any matrices $A$, $B$ and $C$.

Vectorizing equation \eqref{AppGeneralDiffVUp} we find
\be
\nonumber
\frac{\d}{\d t}\vec(\sigma(t))
=(\Omega A \otimes \openone_{2N}
+\openone_{2N} \otimes \Omega A)
\vec(\sigma)+\vec(C).
\ee
We can now apply the same technique which we applied to equations \eqref{AppUnitaryMasterEq} and \eqref{AppGeneralDiffXUp} to find,
\be
\vec(\sigma(t))=M(t)\vec(\sigma(0))+\bm{r}(t)
\ee
where
\begin{align}
M(t)&=\exp\big((\Omega A \otimes \openone_{2N}
+\openone_{2N} \otimes \Omega A) t\big)\\
&\nonumber
=\exp(\Omega A \, t) \otimes \exp(\Omega A \, t)\\
\bm{r}(t)&=\frac{\exp((\Omega A \otimes \Omega A)  t)-\openone_{4N}}{\Omega A \otimes \Omega A}\vec(C).
\end{align}
Unvectorizing this we have
\be
\sigma(t)=(\vec^{-1} M(t)\vec)[\sigma(0)]+\vec^{-1}\big(\bm{r}(t)\big).
\ee
We can simplify this by computing
\begin{align}
&(\vec^{-1} M(t)\vec)[\sigma(0)]\\
&\nonumber
=\vec^{-1}\Big(\big(\exp(\Omega A \, t) \otimes \exp(\Omega A \, t)\big)\vec(\sigma(0))\Big)\\
&\nonumber
=\vec^{-1}\Big(\vec\big(\exp(\Omega A \, t)\sigma(0)\exp(\Omega A \, t)^\intercal\big)\Big)\\
&\nonumber
=\exp(\Omega A \, t)\sigma(0)\exp(\Omega A \, t)^\intercal\\
&\nonumber
=T(t)\sigma(0)T(t)^\intercal
\end{align}
as expected and by defining
\be
R(t)\coloneqq\text{vec}^{-1}\Big(\frac{\exp((\Omega A\otimes\Omega A) \,  t)-\openone_{4N}}{\Omega A\otimes\Omega A} \ \text{vec}(C)\Big).
\ee
This confirms the claim made in \eqref{TdRExplicit}.

\bibliography{references}
\end{document}